\documentclass[11pt,leqno]{article}
\usepackage{a4wide,amsmath,amssymb,amsthm,verbatim,bm}
\newtheorem{theorem}{Theorem}[section]
\newtheorem{lemma}{Lemma}[section]

\newtheorem{corollary}{Corollary}[section]

\newtheorem{remark}{Remark}[section]

\renewcommand{\Re}{\operatorname{Re}}
\renewcommand{\Im}{\operatorname{Im}}
\newcommand{\R}{\mathbb{R}}
\newcommand{\C}{\mathbb{C}}

\newcommand{\N}{\mathbb{N}}

\DeclareMathOperator{\sech}{sech}
\DeclareMathOperator{\spann}{span}
\begin{document}

\title{B\"{a}cklund transformation and $L^2$-stability of NLS solitons}

\author{Tetsu Mizumachi$^{1}$ and Dmitry Pelinovsky$^{2}$ \\
{\small $^{1}$ Faculty of Mathematics, Kyushu University,
Fukuoka 819-0395, Japan} \\
{\small $^{2}$ Department of Mathematics, McMaster
University, Hamilton ON, Canada, L8S 4K1} }
\date{}
\maketitle

\begin{abstract}
Ground states of a $L^2$-subcritical focusing nonlinear Schr\"{o}dinger (NLS)
equation are known to be orbitally stable in the energy class $H^1(\R)$
thanks to its variational characterization.
In this paper, we will show $L^2$-stability of
$1$-solitons to a one-dimensional cubic NLS equation in the sense that
for any initial data which are sufficiently close to a $1$-soliton in $L^2(\R)$,
the solution remains in an $L^2$-neighborhood of a nearby $1$-soliton
solution  for all the time. The proof relies on the B\"{a}cklund
transformation between zero and soliton solutions of this integrable equation.
\end{abstract}
\maketitle

\section{Introduction}
\label{sec:intro}

In this paper, we study the nonlinear Schr\"odinger (NLS) equation
\begin{equation}
\label{NLS-integrable}
iu_t + u_{xx} + 2 |u|^2u= 0\,, \tag{NLS}
\end{equation}
where $u(t,x) : \R \times \R \to \C$. The NLS equation
arises in various areas to describe quasi-monochromatic waves such as laser
beams or capillary gravity waves.
It is well known that \eqref{NLS-integrable} is  well-posed in $L^2$
\cite{T,LP} and in $H^k$ for any $k \in \N$  \cite{GV2,Kato}.
Moreover, solutions of \eqref{NLS-integrable} satisfy conservation laws
for the charge $N$ and Hamiltonian $H$,
\begin{eqnarray}
N(u(t,\cdot)):= \|u(t,\cdot)\|^2_{L^2} = N(u(0,\cdot))\,,\\
H(u(t,\cdot)):= \| \partial_xu(t,\cdot)\|^2_{L^2} - \|u(t,\cdot)\|^4_{L^4} 
= H(u(0,\cdot))\,,
\end{eqnarray}
from which global existence follows in $L^2$ or $H^1$. Note that
\eqref{NLS-integrable} has actually an infinite set of conserved quantities
that resemble norms in $H^k$ for any $k \in \N$ \cite{ZS1}
and these quantities give global existence in $H^k$ for any $k \in \N$.

The NLS equation has a family of solitary waves
(called $1$-solitons) that are written as
\begin{equation}
\label{soliton-general}
u(t,x) = Q_{k,v}(t-t_0,x-x_0)\,,\quad
Q_{k,v}(t,x):= Q_k(x-vt)\, e^{i vx/2+ i(k^2 -v^2/4)t}\,,
\end{equation}
where $Q_k(x)=k\sech(kx)$ and $(k,v,x_0,t_0) \in
\R_+ \times \R \times \R \times \R$ are arbitrary
parameters.

These $1$-solitons play an important role to describe the long-time behavior of
solutions of \eqref{NLS-integrable}. Since $Q_k$ is a minimizer of the functional $H(u)$ restricted on a manifold
$M = \{u \in H^1(\R)\,: \quad \|u\|_{L^2}=\|Q_k\|_{L^2}\}$,
the $1$-soliton (\ref{soliton-general}) is stable in $H^1$ up to
translations in space and time variables (see, e.g., \cite{CL,GSS,We}).
As for orbital stability of $1$-solitons to rougher perturbations,
Colliander {\em et al.} (see \cite{I-team}) show that the $H^s$-norm $(0<s<1)$
of a perturbation to a soliton grows at most polynomially in time
if the initial data is close to the soliton in $H^s(\R)$
($0<s<1$) but not necessarily in $H^1(\R)$.
The result of \cite{I-team} suggests that even for rough initial data for which
the Hamiltonian is not well defined, the $1$-soliton (\ref{soliton-general})
could be stable.

In this paper, we aim to show Lyapunov stability of $1$-solitons in the
$L^2$ class. Our idea is to use the B\"{a}cklund transformation to define
an isomorphism which maps solutions in an $L^2$-neighborhood of
the zero solution to those in an $L^2$-neighborhood of a $1$-soliton
and utilize the $L^2$-stability of the zero solution.

The integrability via the inverse scattering transform method has been
exploited in many details for analysis of spectral stability of
solitary and periodic wave solutions \cite{KSF,Kap}.  It was also used
to analyze orbital stability of dark solitons in the defocusing
version of the NLS equation \cite{Gerard} and to analyze the
long-time asymptotics of solutions of the NLS equation \cite{DZ}.
However, $L^2$-stability of $1$-solitons of (\ref{NLS-integrable})
using the B\"{a}cklund transformation have not been addressed in literature.
In particular, the solvability of the Lax equations to generates the B\"acklund
transformation in the $L^2$-framework is beyond the standard formalism
of the inverse scattering of the NLS equation which requires the
initial data to be in $L^1$, see Lemma 2.1 in \cite{Ablowitz}.

This is not the first time that the integrability is used to prove
stability of solitary waves in the context of other nonlinear
evolution equations.  Merle and Vega \cite{MV} used the Miura
transformation and proved that $1$-solitons of the Korteweg-de Vries
(KdV) equation are stable to $L^2$-perturbations.  The idea was
recently applied by Mizumachi and Tzvetkov \cite{MT} to prove
$L^2$-stability of line solitons of the Kadomtsev-Petviashvili
(KP-II) equation.  The Miura transformation is one of the B\"acklund
transformations which connects solutions of the KdV and the modified KdV
equations. The B\"acklund transformation seems to give a simplified
local coordinate frame which facilitates to observe stability of
solitons. In fact, Mizumachi and Pego \cite{MT} proved asymptotic stability of
Toda lattice solitons by using the B\"acklund transformation to show the
equivalence of linear stability of solitons and that of the zero solution.
Our use of the B\"acklund transformation for the $L^2$-stability result of
NLS solitons is expected to be applicable to other nonlinear evolution
equations associated to the AKNS scheme of inverse scattering.

Now let us introduce our main result on $L^2$-stability of $1$-solitons.
\begin{theorem}
\label{theorem-main}
Let $k>0$ and let $u(t,x)$ be a solution of \eqref{NLS-integrable}
in the class
\begin{equation}
\label{class-solutions}
u\in C(\R;L^2(\R))\cap L^8_{loc}(\R;L^4(\R))\,.
\end{equation}
There exist positive constants $C$ and $\varepsilon$ depending only on
$k$ such that if $\| u(0,\cdot) - Q_k\|_{L^2}<\varepsilon$,
then there exist real constants $k_0$, $v_0$, $t_0$, and $x_0$ such that
\begin{equation}
\label{L2stab-result}
\sup_{t\in\R}\|u(t+t_0,\cdot+x_0) - Q_{k_0,v_0}\|_{L^2}
+ |k_0-k| + |v_0| + |t_0| + |x_0| \le C\|u(0,\cdot)-Q_k\|_{L^2}\,.
\end{equation} 
\end{theorem}
\begin{remark}
Theorem \ref{theorem-main} tells us that solutions of (\ref{NLS-integrable})
which are close initially to a $1$-soliton in the $L^2$-norm remain
close to a nearby $1$-soliton solution for all the time
and the speed, phase, gauge and amplitude parameters of a nearby $1$-soliton
are almost the same as those of the original $1$-soliton.
This makes a contrast with the result of Martel and Merle \cite{MM2}
for the KdV equation that shows that perturbations of $1$-solitons in
$H^1(\R)$ can cause a logarithmic growth of the phase shift thanks to
collisions with infinitely many small solitary waves.
To the best of our knowledge, this is the first result for
the cubic NLS equation in the $L^2$ (or $H^k$, $k \in \N$)
framework which shows that a solution remains close to
a neighborhood of a $1$-soliton for all the time.
\end{remark}
\begin{remark}
Asymptotic stability of solitary waves to a generalized nonlinear
Schr\"odinger equation with a bounded potential in one dimension,
\begin{equation}
\label{NLS-potential}
i u_t+u_{xx} = V(x) u - |u|^{2p} u,
\end{equation}
has been studied by using dispersive decay estimates for solutions to
the linearized equation around solitary waves (see \cite{BS} for $p \geq 4$ and
\cite{Cuc,Mizum} for $p \geq 2$).
However the PDE approach has not resolved yet the asymptotic stability of
solitary waves in the NLS equation (\ref{NLS-potential}) with $p = 1$.
The difficulty comes from the slow decay of solutions in the $L^\infty$ norm
which makes difficult to show convergence of modulation parameters of solitary
waves in time.
\end{remark}

The article is organized as follows. Section~\ref{sec:backlund} reviews the
B\"{a}cklund transformation for the NLS equation.
In Section~\ref{sec:pull-back}, we pull back initial data around a $1$-soliton
to data around the zero solution by solving the B\"acklund transformation
at $t=0$. When we solve the B\"acklund transformation around a $1$-soliton
solution at $t=0$, the parameters which describe the amplitude, the velocity,
and the phase shifts of the time and space variables of the largest soliton 
in the solution are uniquely determined. This shows one of the difference between
our approach and the method based on the modulation theory
(see, e.g., \cite{BS,Cuc,Mizum}), where convergence of varying parameters in time 
is achieved using decay estimates of the dispersive part of the solution.

In Section~\ref{sec:backlund-time-evolution}, we prove that the B\"acklund
transformation defines a continuous mapping from an $L^2$-neighborhood of
the origin to an $L^2$-neighborhood of a $1$-soliton and that the B\"acklund
transformation connects solutions around $1$-solitons and solutions around the
zero solution for all the time if initial data are smooth.
Thanks to the $L^2$-conservation law of the NLS equation, the zero solution is
stable in $L^2$ and we conclude that if a perturbation to initial data is small
in $L^2$, then a solution stays in $L^2$-neighborhood of the $1$-soliton
obtained in Section~\ref{sec:pull-back}. Section 5 concludes the article with
discussion of open problems.

\section{B\"{a}cklund transformation for the NLS equation}
\label{sec:backlund}
We recall the B\"{a}cklund transformation between two different
solutions $q(t,x)$ and $Q(t,x)$ of
(\ref{NLS-integrable}). This transformation was found in two different
but equivalent forms \cite{Chen,Wadati}.

The NLS equation is a solvability condition of the
Lax operator system
\begin{equation}
\label{Lax-1}
\partial_x \left[ \begin{array}{c} \psi_1 \\ \psi_2 \end{array} \right] =
\left[ \begin{array}{cc} \eta & q \\ -\bar{q} &  - \eta \end{array} \right]
\left[ \begin{array}{c} \psi_1 \\ \psi_2 \end{array} \right]
\end{equation}
and
\begin{equation}
\label{Lax-2}
\partial_t \left[ \begin{array}{c} \psi_1 \\ \psi_2 \end{array} \right] = i
\left[ \begin{array}{cc} 2 \eta^2 + |q|^2 &  \partial_x q + 2 \eta q \\
\partial_x \bar{q} - 2 \eta \bar{q} &  - 2 \eta^2 - |q|^2 \end{array} \right]
\left[ \begin{array}{c} \psi_1 \\ \psi_2 \end{array} \right],
\end{equation}
where parameter $\eta$ is $(t,x)$-independent.

Using the variable
$$
\gamma = \frac{\psi_1}{\psi_2},
$$
we obtain the Riccati equations for the NLS equation
\begin{equation}
\label{Riccati}
\left\{ \begin{array}{l} \partial_x \gamma = 2\eta\gamma + q
 +\bar{q} \gamma^2, \\
\partial_t \gamma = i (4\eta^2+ 2 |q|^2) \gamma + i( \partial_x q + 2\eta q)
-i (\partial_x \bar{q} - 2\eta \bar{q}) \gamma^2\,.
\end{array} \right.
\end{equation}
A new solution $Q(t,x)$ of the same equation
(\ref{NLS-integrable}) is obtained from the old solution $q(t,x)$ and
the solution $\gamma(t,x)$ of the Riccati equations (\ref{Riccati})
(or equivalently, from the solution $\psi_1(t,x)$ and $\psi_2(t,x)$ of the Lax
equations (\ref{Lax-1})--(\ref{Lax-2})) by
\begin{equation}
Q + q = \frac{-4\Re(\eta) \gamma}{1 + |\gamma|^2}
=\frac{-4\Re(\eta) \psi_1 \bar{\psi}_2}{|\psi_1|^2 + |\psi_2|^2}.
\label{new-solution}
\end{equation}

The new solution $Q$ appears as the potential in the same Riccati equations (\ref{Riccati})
for $\Gamma$ and in the same Lax equations (\ref{Lax-1})--(\ref{Lax-2}) for
$\Psi_1$ and $\Psi_2$ if
\begin{equation}
\label{correspondence}
\Gamma = \frac{1}{\bar{\gamma}}, \quad \Psi_1 = \frac{\bar{\psi}_2}{|\psi_1|^2 + |\psi_2|^2}, \quad
\Psi_2 = \frac{\bar{\psi}_1}{|\psi_1|^2 + |\psi_2|^2}.
\end{equation}

As a simple example, we can start from the zero solution $q(x,t) \equiv 0$ and
assume that $k=2\eta$ is a real positive number.
Equations (\ref{Lax-1})--(\ref{new-solution})
give a soliton solution
\begin{equation}
\label{soliton}
Q(t,x) = Q_k(x) e^{i k^2 t}\,, \quad Q_k(x) := k \; \sech(kx),
\end{equation}
if
\begin{equation}
\label{soliton-transformation}
\psi_1 = e^{(kx+ik^2t)/2}, \quad \psi_2 = -e^{-(kx+ik^2t)/2}, \quad \gamma = -e^{kx+ik^2t}
\end{equation}
or equivalently,
\begin{equation}
\label{soliton-transformation-inverse}
\Psi_1 = -\frac{e^{-(kx-ik^2t)/2}}{2 \cosh(kx)},
\quad \Psi_2 = \frac{e^{(kx-ik^2t)/2}}{2 \cosh(kx)}, \quad
\Gamma = -e^{-kx+ik^2t}.
\end{equation}
Compared to a general family of $1$-solitons (\ref{soliton-general}),
solution (\ref{soliton}) is centered at $x = 0$ and has zero velocity and zero phase.

\begin{remark}
\label{remark-1}
If we eliminate the variable $\gamma$ from equation (\ref{new-solution}) and
close the system of equations (\ref{Riccati}) for the new and old solutions $Q$ and $q$,
then $\gamma$ satisfies a quadratic equation that has two roots
\begin{equation}
\label{two-roots}
\gamma = -\frac{k \pm \sqrt{k^2 - |Q + q|^2}}{\bar{Q} + \bar{q}}.
\end{equation}
This form of the B\"acklund transformation was considered in \cite{Chen,Wadati}. Unfortunately, the explicit solution
(\ref{soliton}) and (\ref{soliton-transformation}) show that the upper root in (\ref{two-roots})
is taken for $x > 0$ and the lower root in (\ref{two-roots})
is taken for $x < 0$ with a weak singularity at $x = 0$.
\end{remark}

\begin{remark}
\label{remark-uniqueness}
General solutions of the Lax equations (\ref{Lax-1})--(\ref{Lax-2}) for $q=0$
and $\eta=(k+iv)/2$ with $(k,v) \in \R^2$ are given by
$$
\psi_1(t+t_0,x+x_0) = e^{(k(x-2vt)+i\omega t+ivx)/2}, \quad
\psi_2(t+t_0,x+x_0) = -e^{-(k(x-2vt)+i\omega t+ivx)/2}\,,
$$
where $(x_0,t_0) \in \R^2$ are arbitrary parameters for the soliton position
and phase, and $\omega=k^2-v^2$.
\end{remark}

\section{From a $1$-soliton to the zero solution at $t=0$}
\label{sec:pull-back}
In this section, we will pull back solutions around a 1-soliton to those
around the zero solution by using the B\"acklund transformation at time $t=0$.

Let us define $q(0,x)$ by the B\"acklund transformation
\begin{equation}
Q+q = \frac{-4 \Re(\eta) \Psi_1 \overline{\Psi_2}}{|\Psi_1|^2 + |\Psi_2|^2},
\label{new-solution-converse}
\end{equation}
associated to  solutions of the Lax equation
\begin{equation}
\label{Lax-1-converse}
\partial_x \left[ \begin{array}{c} \Psi_1 \\ \Psi_2 \end{array} \right] =
\left[ \begin{array}{cc} \eta & Q \\ -\bar{Q} &  - \eta
\end{array} \right]
\left[ \begin{array}{c} \Psi_1 \\ \Psi_2 \end{array} \right].
\end{equation}
When $\eta=\frac12$ and $Q(x) = Q_1(x) \equiv \sech(x)$, the spectral problem
(\ref{Lax-1-converse}) has a fundamental system
$\{\mathbf{\Psi}_1(x),\mathbf{\Psi}_2(x)\}$, where
\begin{equation}
\label{fund-sol}
\mathbf{\Psi}_1(x)=\left[ \begin{array}{c} -e^{-x/2} \\ e^{x/2} \end{array} \right]
 \; \sech(x)\,, \quad
\mathbf{\Psi}_2=\left[
\begin{array}{c} (e^x + 2(1+x) e^{-x}) e^{x/2} \\ (e^{-x} - 2 x e^x) e^{-x/2}
\end{array} \right] \; \sech(x)\,.
\end{equation}
We obtain $q = 0$ when the first solution $\mathbf{\Psi}_1$ is used in
the B\"{a}cklund transformation \eqref{new-solution-converse}
with $\eta=\frac12$ and
\begin{equation}
q(x) = \frac{2x e^{2x} + (4x^2 + 4x - 1) - 2 x(1+x) e^{-2x}}{\cosh(3x) + 4(1 + x + x^2) \cosh(x)} - \sech(x)
\end{equation}
when the second solution $\mathbf{\Psi}_2$ is used in
\eqref{new-solution-converse} with $\eta=\frac12$.
The latter solution corresponds to the weak (logarithmic in time)
scattering of two nearly identical solitons. This interaction between two solitons
was studied by Zakharov and Shabat \cite{ZS} shortly after the integrability
of the NLS equation was discovered by the same authors \cite{ZS1}.
We are interested in the decaying solution of the spectral problem
(\ref{Lax-1-converse}), which corresponds to the eigenvector for a
simple isolated eigenvalue $\eta = \frac{1}{2}$ associated
to the potential $Q_1(x) = \sech(x)$.

Let us recall the Pauli matrices
\begin{gather*}
\sigma_1=\begin{bmatrix}0&1\\1&0\end{bmatrix}\,,\quad
\sigma_2=\begin{bmatrix}0&-i\\i&0\end{bmatrix}\,,\quad
\sigma_3=\begin{bmatrix}1&0\\0&-1\end{bmatrix}\,.
\end{gather*}
The spectral problem \eqref{Lax-1-converse} is equivalent to
an eigenvalue problem
\begin{equation}
\label{eq:evp}
(L-M(S))\mathbf{\Psi}=\lambda \mathbf{\Psi},
\end{equation}
where $\lambda=\eta-\frac12$, $S=Q-Q_1$,
$$
L := \left[ \begin{array}{cc} \partial_x -\frac{1}{2} & - Q_1 \\
- Q_1 & -\partial_x - \frac{1}{2} \end{array} \right] = \sigma_3 \partial_x  -\frac{1}2 I
- Q_1\sigma_1 \equiv L_0 - Q_1\sigma_1,
$$
and
$$
M(S) := \left[ \begin{array}{cc} 0 & S \\ \bar{S} &  0 \end{array} \right] = \sigma_1  \Re(S) - \sigma_2 \Im(S).
$$

We consider $L$ as a closed operator on $L^2(\R;\C^2)$ whose domain
is  $H^1(\R;\C^2)$.
If $S = 0$, then $\lambda=0$ is an eigenvalue of \eqref{eq:evp}
whose eigenspace is spanned by $\mathbf{\Psi}_1$.
Since
$$\left(M(Q_1)L_0^{-1}\begin{bmatrix} f_1 \\ f_2 \end{bmatrix}\right)(x)
=-Q_1(x)\begin{bmatrix} \int^x_{-\infty} e^{-(x-y)/2}f_2(y)dy
\\ \int_x^{\infty} e^{(x-y)/2}f_1(y)dy\end{bmatrix}\,,
$$
we see that $M(Q_1)L_0^{-1}$ is Hilbert-Schmidt and thus
a compact operator on $L^2(\R;\C^2)$. Thus by Weyl's essential spectrum
theorem, we have
$\sigma_c(L) =\sigma(L_0)=\{ -\frac{1}{2} + i k, \;\; k \in \R \}$  and
the zero eigenvalue is bounded away from the rest of the spectrum of $L$.
Thus for small $S$, we will see that the eigenvalue problem \eqref{eq:evp} has a simple
eigenvalue near $0$.

\begin{lemma}
\label{lemma-1-converse}
There exist positive constants $C$, $\varepsilon$ and real constants
$k$, $v$ such that if $\| Q - Q_1 \|_{L^2} \leq \varepsilon$, then there exist
a solution $\mathbf{\Psi}={}^t(\Psi_1,\Psi_2)\in H^1(\R;\C^2)$ of the system
\eqref{Lax-1-converse} with $\eta=(k+iv)/2$ such that
\begin{equation}
\label{bound-solution}
|k-1|+|v|+\|\mathbf{\Psi}-\mathbf{\Psi}_1 \|_{L^\infty} \leq C\|Q-Q_1\|_{L^2} \,.
\end{equation}
\end{lemma}

\begin{proof}
We will prove Lemma~\ref{lemma-1-converse} by the Lyapunov-Schmidt method.
Let us write $Q = Q_1 +S$ and
\begin{equation}
\label{equation-0}
\mathbf{\Psi} = \mathbf{\Psi}_1 + \mathbf{\Phi}, \quad \langle
\mathbf{\Psi}_1, \mathbf{\Phi} \rangle_{L^2} = 0.
\end{equation}
Let $P$ be a spectral projection associated with $L$ on $L^2(\R;\C^2)$,
or explicitly,
$$
P\mathbf{u}=\mathbf{u}-\frac{1}{4}\langle \mathbf{u},\mathbf{\Theta}
\rangle_{L^2} \mathbf{\Psi}_1\,,\quad \mathbf{\Theta}(x)=
\begin{bmatrix} -e^{x/2} \\ e^{-x/2} \end{bmatrix}\, \sech(x).
$$
Note that $\ker(L)=\spann\{\mathbf{\Psi}_1\}$ and
$\ker(L^*)=\spann\{\mathbf{\Theta}\}$.
The system (\ref{Lax-1-converse}) can be rewritten into the
block-diagonal form
\begin{equation}
\label{equation-2}
L \mathbf{\Phi} =P\left[(\lambda I+ M(S))
(\mathbf{\Psi}_1 + \mathbf{\Phi})\right],
\end{equation}
and
\begin{equation}
\label{equation-3}
\langle \mathbf{\Theta}, (\lambda I+M(S))(\mathbf{\Psi} + \mathbf{\Phi})) \rangle_{L^2} = 0.
\end{equation}
Since $L_0$ is a closed operator on $L^2(\R;\C^2)$ with
$\operatorname{Range}(L_0)=L^2(\R;\C^2)$ and $M(Q_1)L_0^{-1}$ is a compact
operator on $L^2(\R;\C^2)$, we see that $L$ is Fredholm and
$$
\operatorname{Range}(L)=\{\mathbf{\Phi}\in L^2(\R;\C^2): \langle\mathbf{\Phi},
\mathbf{\Theta}\rangle_{L^2}=0\}.
$$
Thus we can define $L^{-1}$ as a bounded operator
$$L^{-1}:L^2(\R;\C^2)\cap {}^\perp\ker(L^*)\to H^1(\R;\C^2)\cap {}^\perp\ker(L).$$
\par
If $S \in L^2(\R)$ and $\lambda \in \C$ are sufficiently small,
there exists a unique solution $\mathbf{\Phi}\in H^1(\R^2;\C^2)$ of
\eqref{equation-2} such that
\begin{equation}
\label{equation-4}
\|\mathbf{\Phi} \|_{H^1 \times H^1} \leq C(\|S\|_{L^2}+|\lambda|)\,,
\end{equation}
where $C$ is a constant that does not depend on $S$ and $\lambda$.
On the other hand, equation (\ref{equation-3}) can be written in the form
\begin{align*}
& \lambda \left(4+\int_{\R} \sech(x)
\left[ -e^{x/2} \Phi_1(x) + e^{-x/2} \Phi_2(x)\right] dx \right)
\\ =&
2 \langle Q_1,\Re(S) \rangle_{L^2} - 2 i \langle \partial_x Q_1,
\Im(S) \rangle_{L^2} - \int_{\R} \sech(x) \left[e^{x/2} S(x) \Phi_2(x)
+e^{-x/2} \overline{S(x)} \Phi_1(x) \right] dx
\end{align*}
In view of the bound (\ref{equation-4}), the latter equation gives
\begin{equation}
\label{equation-5}
\exists C > 0 : \quad
\left| \lambda - \frac{1}{2} \langle Q_1, \Re(S) \rangle_{L^2}
+\frac{i}{2} \langle \partial_x Q_1, \Im(S) \rangle_{L^2} \right|
\leq C \| S \|^2_{L^2}\,,
\end{equation}
which concludes the proof of Lemma \ref{lemma-1-converse} since
$\lambda = \eta -\frac{1}{2}$ and $S = Q - Q_1$.
\end{proof}

\begin{remark}
\label{remark-constraints-S}
If the eigenvalue $\eta$ is forced to stay at $\frac{1}{2}$, constraints on $S(x)$ need to be enforced,
which are given at the leading order by
\begin{equation}
\label{equation-orthog}
\langle Q_1,\Re(S) \rangle_{L^2} = 0, \quad
\langle \partial_x Q_1, \Im(S) \rangle_{L^2} = 0.
\end{equation}
\end{remark}

Constraints (\ref{equation-orthog}) are nothing but the symplectic
orthogonality conditions to the eigenvectors of the linearized
time-evolution problem that correspond to the zero eigenvalue induced
by the gauge and translational symmetries of the NLS equation. The
symplectic orthogonality conditions were used in \cite{Cuc,Mizum} to
derive modulation equations for varying parameters of the solitary
wave and to prove its asymptotic stability in the time evolution of
the generalized NLS equation (\ref{NLS-potential}).

Let us generalize the symplectic orthogonal conditions (\ref{equation-orthog})
and decompose $Q$ into a sum of all four secular modes and the residual part.
This decomposition is standard and follows from the implicit function theorem
arguments (see, e.g.,
\cite{Cuc,Mizum}).

\begin{lemma}
\label{lem:Qdecompose}
There exist positive constants $C$, $\varepsilon$ and real constants
$\alpha$, $\beta$, $\theta$, $\gamma$ such that if $\|Q-Q_1\|_{L^2} \leq \varepsilon$,
then $Q$ can be represented by
\begin{equation}
\label{eq:Qdecomp1}
e^{-i(vx+\theta)}Q(\cdot+\gamma)=Q_k+i\alpha xQ_k+\beta\partial_kQ_k+S\,,
\end{equation}
with
\begin{equation}
\label{eq:Qdecomp2}
\langle Q_k,  \Re(S) \rangle_{L^2} = \langle \partial_x Q_k, \Im(S) \rangle_{L^2}
= \langle xQ_k,  \Re(S) \rangle_{L^2} = \langle \partial_k Q_k, \Im(S)
\rangle_{L^2}=0\,
\end{equation}
and
\begin{equation}
\label{eq:Qdecomp3} |\alpha|+|\beta|+|\theta|+|\gamma|+\|S\|_{L^2}\le
C\|Q-Q_1\|_{L^2}\,,
\end{equation}
where $k$ and $v$ are real constants given in Lemma~\ref{lemma-1-converse}.
\end{lemma}

In order to estimate the $L^2$-norm of $q$ defined by the B\"{a}cklund transformation
\eqref{new-solution-converse}, we need to investigate solutions to the
system \eqref{Lax-1-converse}.

\begin{lemma}
\label{lem:growth-rate}
There exist  positive constants $C$ and $\varepsilon$ such that
if $\|Q-Q_1\|_{L^2} \leq \varepsilon$, then an $H^1$-solution of the system
\eqref{Lax-1-converse} with $\eta=(k+iv)/2$ determined in Lemma
\ref{lemma-1-converse} satisfies
\begin{gather}
\label{eq:Psidecomp1}
\mathbf{\Psi}(x+\gamma) = \sech(kx)
e^{\frac{i}{2}(vx+\theta)\sigma_3}\begin{bmatrix}e^{-kx/2}(-1+r_{11}(x))+e^{kx/2}r_{12}(x)
\\  e^{-kx/2}r_{21}(x)+e^{kx/2}(1+r_{22}(x))\end{bmatrix}\,,\\
\label{eq:Psidecomp2}
\|r_{11}\|_{L^\infty}+\|r_{12}\|_{L^2\cap L^\infty}+\|r_{21}\|_{L^2\cap L^\infty}
+\|r_{22}\|_{L^\infty}\le C \|Q-Q_1\|_{L^2}\,,
\end{gather}
where $\gamma$ and $\theta$ are constants determined in
Lemma~\ref{lem:Qdecompose}. Moreover if  $Q\in H^n(\R)$ $(n\in\N)$
in addition, then 
\begin{equation}
  \label{eq:Psidecomp2'}
\|\partial_x^m r_{11}\|_{L^\infty} +\|\partial_x^ mr_{12}\|_{L^2\cap L^\infty}
+\|\partial_x^m r_{21}\|_{L^2\cap L^\infty} +\|\partial_x^mr_{22}\|_{L^\infty}
\le  C'(\|Q-Q_1\|_{H^m}+\|Q-Q_1\|_{H^m}^m)
\end{equation}
for $0\le m\le n$, where $C'$ is a positive constant depending only on $n$.
\end{lemma}

Lemma~\ref{lem:growth-rate} will be proven in the end of this section.
Assuming Lemma~\ref{lem:growth-rate}, we will prove that the B\"acklund
transformation maps initial data around a $1$-soliton to those around the 
zero solution.

\begin{lemma}
\label{lemma-2-converse}
There exist positive constants  $C$ and $\varepsilon$ satisfying the following:
Let $Q\in H^3(\R)$ and $\|Q-Q_1\|_{L^2} \leq \varepsilon$ and let
$\mathbf{\Psi}$ be an $H^1$-solution of the
system \eqref{Lax-1-converse} with $\eta = (k+iv)/2$ determined in
Lemma~\ref{lemma-1-converse}. Suppose 
$$
q := -Q- \frac{2k\Psi_1 \overline{\Psi_2}}{|\Psi_1|^2 + |\Psi_2|^2}\,.
$$
Then $q\in H^3(\R)$ and $\|q \|_{L^2} \leq C \| Q - Q_1 \|_{L^2}$.
\end{lemma}

\begin{proof}
By \eqref{eq:Psidecomp1} and \eqref{eq:Psidecomp2},
we have
\begin{equation}
\label{eq:l21}
\begin{split}
-\frac{2k\Psi_1\overline{\Psi_2}}{|\Psi_1|^2+|\Psi_2|^2} = 2ke^{i(vx(x-\gamma)+\theta)}
\frac{1+\varepsilon_1(x)+e^{kx(x-\gamma)}\varepsilon_2(x)+e^{-k(x-\gamma)}\varepsilon_3(x)}
{e^{kx(x-\gamma)}(1+\varepsilon_4(x))+\varepsilon_5(x)+e^{-kx(x-\gamma)}(1+\varepsilon_6(x))}\,,
\end{split}
\end{equation}
where
\begin{align*}
& \varepsilon_1=\bar{r}_{22}-r_{11}-r_{12}\bar{r}_{21}-r_{11}\bar{r}_{22}\,, \\
& \varepsilon_2=-(1+\bar{r}_{22})r_{12}\,,\\ 
& \varepsilon_3=\bar{r}_{21}(1-r_{11})\,,\\
& \varepsilon_4=2\Re(r_{22}) + |r_{22}|^2+|r_{12}|^2\,, \\
& \varepsilon_5=-2\Re(\bar{r}_{12}(1-r_{11})) + 2 \Re(\bar{r}_{21}(1+r_{22}))\,, \\
& \varepsilon_6=-2\Re(r_{11}) +|r_{11}|^2+|r_{21}|^2\,.
\end{align*}

Lemmas~\ref{lemma-1-converse}, \ref{lem:Qdecompose} and \ref{lem:growth-rate}
imply that
$$
|k-1|+|v|+|\theta|+|\gamma| \lesssim \|Q-Q_1\|_{L^2}\,
$$
and
$$
\|\varepsilon_1\|_{L^\infty} + \|\varepsilon_2\|_{L^2 \cap L^{\infty}} + \|\varepsilon_3\|_{L^2 \cap L^{\infty}} +
\|\varepsilon_4\|_{L^\infty} + \|\varepsilon_5\|_{L^2 \cap L^{\infty}} + \|\varepsilon_6\|_{L^\infty}
\lesssim \|Q-Q_1\|_{L^2}\,,
$$
where notation $A \lesssim B$ is used to say that there is a positive constant
$C$ such that $A \leq C B$. 

Combining the above bounds with the expansion, 
\begin{align*}
& \frac{1+\varepsilon_1(x)+e^{kx(x-\gamma)}\varepsilon_2(x)+e^{-k(x-\gamma)}\varepsilon_3(x)}
{e^{kx(x-\gamma)}(1+\varepsilon_4(x))+\varepsilon_5(x)+e^{-kx(x-\gamma)}(1+\varepsilon_6(x))}
\\=& \frac{1+\varepsilon_1(x)}
{e^{kx(x-\gamma)}(1+\varepsilon_4(x))+\varepsilon_5(x)+e^{-kx(x-\gamma)}(1+\varepsilon_6(x))}
+ {\cal O}(|\varepsilon_2(x)|+|\varepsilon_3(x)|)
\\ =& \frac12\sech\left(k(x-\gamma)\right)
\left(1 + {\cal O}(|\varepsilon_1(x)|+|\varepsilon_4(x)|+|\varepsilon_5(x)|+|\varepsilon_6(x)|)\right) + {\cal O}(|\varepsilon_2(x)|+|\varepsilon_3(x)|)\,,
\end{align*}
we get
\begin{equation}
\label{eq:*1}
\exists C > 0 : \quad
\left\|\frac{2k\Psi_1\overline{\Psi_2}}{|\Psi_1|^2+|\Psi_2|^2}
  +Q_1\right\|_{L^2}\le C\|Q-Q_1\|_{L^2}.
\end{equation}
Thus by \eqref{new-solution-converse} and \eqref{eq:*1},
\begin{align*}
\|q\|_{L^2}\le \|Q-Q_1\|_{L^2}+
\left\|\frac{2k\Psi_1\overline{\Psi_2}}{|\Psi_1|^2+|\Psi_2|^2}
+Q_1\right\|_{L^2} \le (C+1)\|Q-Q_1\|_{L^2}\,.
\end{align*}
If $Q\in H^3(\R)$ in addition, then it follows from
\eqref{new-solution-converse}, \eqref{eq:Psidecomp2'} and
\eqref{eq:l21} that $q\in H^3(\R)$.
\end{proof}

\begin{corollary}
Under conditions of Lemma~\ref{lemma-2-converse}, let
$$
\psi_1=\frac{\overline{\Psi_2}}{|\Psi_1|^2 + |\Psi_2|^2}\,,\quad
\psi_2=\frac{\Psi_1}{|\Psi_1|^2 + |\Psi_2|^2}\,.
$$
Then $(\psi_1,\psi_2)$ are $C^2$-functions satisfying
\eqref{Lax-1}.
\end{corollary}

\begin{proof}
Lemma~\ref{lem:growth-rate} implies that $\psi_1$ and
$\psi_2$ are $C^2$-functions. By a direct substitution, we see that
$(\psi_1,\psi_2)$ is a solution of the system \eqref{Lax-1}.
\end{proof}

\begin{remark} \label{remark-change}
Using the change of variables
\begin{eqnarray*}
\Psi'_1(y) & = & e^{-\frac{i}2(vx+\theta)}\Psi_1(x+\gamma)\,,\\
\Psi'_2(y) & = & e^{\frac{i}2(vx+\theta)}\Psi_2(x+\gamma)\,, \\
Q'(y) & = & k^{-1}e^{-i(vx+\theta)}Q(x+\gamma)\,,
\end{eqnarray*}
where $y = kx$, we can translate the system \eqref{Lax-1-converse} with $\eta = (k+iv)/2$ into
$$
\partial_y \begin{bmatrix}\Psi'_1 \\ \Psi'_2\end{bmatrix}
= \begin{bmatrix}\frac 12 & Q' \\ -\overline{Q'} &
-\frac 12 \end{bmatrix}
\begin{bmatrix}\Psi'_1 \\ \Psi'_2\end{bmatrix}\,.
$$
Therefore, we will assume $k=1$ and $v=\gamma=\theta=0$
in \eqref{eq:Qdecomp1} and \eqref{eq:Qdecomp2} for the sake of simplicity.
\end{remark}

Next we will give an estimate of solutions to the linear inhomogeneous equation
\begin{equation}
  \label{eq:luf}
L\mathbf{u}=\mathbf{f} \,.
\end{equation}
To prove Lemma~\ref{lem:growth-rate}, we introduce Banach spaces
$X=X_1\times X_2$ and $Y=Y_1\times Y_2$ such that
for $\mathbf{u}={}^t(u_1\,,\,u_2)\in X$ and $\mathbf{f}={}^t(f_1\,,\,f_2)\in Y$,
we have
\begin{eqnarray*}
\|\mathbf{u}\|_X = \|u_1\|_{X_1}+\|u_2\|_{X_2}\,,\quad
\|\mathbf{f}\|_Y = \|f_1\|_{Y_1}+\|f_2\|_{Y_2}\,,
\end{eqnarray*}
equipped with the norms
\begin{eqnarray*}
\|u_1\|_{X_1} & := & \inf_{u_1=v_1+w_1} \left( \|e^{x/2}\cosh(x)\; v_1\|_{L^\infty}
+\|e^{-x/2}\cosh(x)\; w_1\|_{L^2\cap L^\infty} \right)\,,
\\ \|u_2\|_{X_2} & := & \inf_{u_2=v_2+w_2} \left( \|e^{-x/2}\cosh(x)\; v_2\|_{L^\infty}
+\|e^{x/2}\cosh(x)\; w_2\|_{L^2\cap L^\infty} \right)\,
\end{eqnarray*}
and
\begin{eqnarray*}
\|f_1\|_{Y_1} & := & \inf_{f_1=g_1+h_1} \left(\|e^{-x/2}\cosh(x)\; g_1\|_{L^2}
+\|e^{x/2}\cosh(x)\; h_1\|_{L^1\cap L^2} \right)\,, \\
 \|f_2\|_{Y_2} & := & \inf_{f_2=g_2+h_2} \left( \|e^{x/2}\cosh(x)\; g_2\|_{L^2}
+\|e^{-x/2}\cosh(x)\; h_2\|_{L^1\cap L^2} \right)\,.
\end{eqnarray*}

\begin{lemma}
\label{lem:l1-est}
Let $\mathbf{f}={}^t(f_1,f_2)\in Y\cap {}^\perp\ker(L^*)$ and
let $\mathbf{u}$ be a solution of the system \eqref{eq:luf} such that
$\mathbf{u} \perp \ker(L)$. Then, there is an $\mathbf{f}$-independent
constant $C > 0$ such that
$\|\mathbf{u}\|_{X}\le C\|\mathbf{f}\|_{Y}$.
\end{lemma}

\begin{remark}
For an arbitrary $\mathbf{f}\in L^2(\R;\C^2)\cap {}^\perp\ker(L^*)$,
an $H^1$-solution $\mathbf{u}$ of the system \eqref{eq:luf} does not necessarily decay as fast
as its fundamental solution. However, since the potential matrix $M(S)$ in
\eqref{eq:evp} is off-diagonal, solutions have a better decay property, according to 
the norm in $X$.
\end{remark}

To prove Lemma~\ref{lem:l1-est}, we will use an explicit formula of
$L^{-1}\mathbf{f}$.

\begin{lemma}
\label{lem:resolvent}
For any $\mathbf{f}={}^t(f_1,f_2)\in L^2(\R;\C^2)\cap {}^\perp\ker(L^*)$,
there exists a unique solution
$\mathbf{u}\in H^1(\R;\C^2)\cap {}^\perp\ker(L)$ of the system \eqref{eq:luf} that
can be written as
\begin{equation}
  \label{eq:resolve}
  \begin{split}
\mathbf{u}(x)= & \zeta(\mathbf{f}) \mathbf{\Psi}_1(x)+
\frac{1}{4} \mathbf{\Psi}_1(x) \int_x^\infty
e^{y/2}(e^{-2y}-2y)\sech(y) f_1(y)dy \\ &
-\frac{1}{4} \mathbf{\Psi}_1(x) \int_{-\infty}^xe^{-y/2}(e^{2y}+2+2y)\sech(y) f_2(y)dy
+\frac{1}{4} \mathbf{\Psi}_2(x) \int^{\pm\infty}_x \mathbf{f}(y)\cdot\mathbf{\Theta}(y)dy\,,
  \end{split}
\end{equation}
where $\zeta(\mathbf{f})$ is continuous linear functional on
$L^2$.
\end{lemma}

\begin{remark}
\label{rem:cancel}
If $\langle \mathbf{f},\mathbf{\Theta}\rangle_{L^2}=0$, then
\begin{equation}
  \label{eq:cancel}
\int_x^\infty\mathbf{f}(y)\cdot\mathbf{\Theta}(y)dy=-\int_{-\infty}^x\mathbf{f}(y)\cdot
\mathbf{\Theta}(y)dy\,.
\end{equation}
\end{remark}

\noindent\textbf{Proof of Lemma~\ref{lem:resolvent}.}
Since $L:H^1(\R;\C^2)\to L^2(\R;\C^2)$ is a Fredholm operator, the equation
\eqref{eq:luf} has a solution in $L^2(\R;\C^2)$ if $\mathbf{f}$ is orthogonal
to $\ker(L^*)=\spann\{\mathbf{\Theta}\}$.

Using a fundamental matrix $U(x) = [\mathbf{\Psi}_1(x),\mathbf{\Psi}_2(x)]$ of
$$\partial_x\mathbf{\Psi}=\begin{bmatrix}\frac12 & Q_1 \\ -Q_1 & -\frac12\end{bmatrix}
\mathbf{\Psi},$$
we rewrite $L\mathbf{u}=\mathbf{f}$ as
\begin{align*}
\frac{d}{dx}(U(x)^{-1}\mathbf{u}) = U(x)^{-1}\sigma_3 \mathbf{f}
= \frac{-1}{4} \sech(x)
\begin{bmatrix}e^{x/2}(e^{-2x}-2x) & e^{-x/2}(e^{2x}+2x+2)\\
-e^{x/2} & e^{-x/2}\end{bmatrix}
\begin{bmatrix}f_1(x) \\ f_2(x)\end{bmatrix}\,.
\end{align*}
Thus we have $$\mathbf{u}(x)= U(x)\mathbf{c}-\frac{1}{4}U(x)\mathbf{g}(x)\,,$$
where $\mathbf{c}$ is a constant vector, $\mathbf{g}(x)={}^t(g_1(x)\,,\,g_2(x))$
and
\begin{align*}
g_{1}(x)=&  \int_{x_1}^x e^{y/2}(e^{-2y}-2y) \sech(y) f_1(y)dy +
\int_{x_2}^x e^{-y/2}(e^{2y}+2y+2) \sech(y) f_2(y)dy\,,
\\ g_{2}(x)=& -\int_{x_3}^x e^{y/2}\sech(y) f_1(y)dy
+\int_{x_4}^x e^{-y/2}\sech(y) f_2(y)dy\,.
\end{align*}
Note that $x_1$, $x_2$, $x_3$, and $x_4$ can be chosen freely.
To let $\mathbf{u}\in L^2(\R;\C^2)$, we put
$x_1=\infty$, $x_2=-\infty$, $x_3=x_4=\pm\infty$ and
$\mathbf{c}={}^t(\zeta\,,\,0)$ and obtain \eqref{eq:resolve}.
\par
Next we will show that $\zeta(\mathbf{f})$ is continuous on $L^2$.
Since $|\mathbf{\Psi}_1(x)|\lesssim e^{-|x|/2}$ for all $x \in \R$,
\begin{equation*}
  \begin{split}
& \left\|\mathbf{\Psi}_1(x)
\int_x^\infty e^{y/2}(e^{-2y}-2y)\sech(y) f_1(y)dy \right\|_{L^2}
\\ \lesssim &
\left\|\mathbf{\Psi}_1(x) \int_x^\infty e^{-3y/2}\sech(y) f_1(y)dy \right\|_{L^2}
+\left\|\mathbf{\Psi}_1(x) \int_x^\infty y e^{y/2}\sech(y) f_1(y)dy \right\|_{L^2}
\\ \lesssim &
\left\|\int_x^\infty e^{(x-y)/2} |f_1(y)|dy \right\|_{L^2}
+\left\|e^{-|x|/2}
\int_x^\infty e^{-|y|/2} |yf_1(y)|dy \right\|_{L^2} \lesssim \|\mathbf{f}\|_{L^2}\,.
  \end{split}
\end{equation*}
Similarly, we have
$$
\left\|\mathbf{\Psi}_1(x) \int_{-\infty}^xe^{-y/2}(e^{2y}+2+2y)\sech(y) f_2(y)dy
\right\|_{L^2} \lesssim  \|\mathbf{f}\|_{L^2}\,.
$$
Using Remark~\ref{rem:cancel} and the fact that $|\mathbf{\Psi}_2(x)|\lesssim e^{|x|/2}$ and
$|\mathbf{\Theta}(x)|\lesssim e^{-|x|/2}$ for all $x \in \R$, we have
\begin{equation*}
  \begin{split}
& \left\|\mathbf{\Psi}_2(x)\int^{\pm\infty}_x \mathbf{f}(y)\cdot\mathbf{\Theta}(y)dy
\right\|_{L^2} \\ \lesssim &
\left\|\int^\infty_x e^{(x-y)/2}|\mathbf{f}(y)|dy\right\|_{L^2(0,\infty)}
+\left\|\int_{-\infty}^x e^{-(x-y)/2}|\mathbf{f}(y)|dy\right\|_{L^2(-\infty,0)}
\lesssim \|\mathbf{f}\|_{L^2}\,.
  \end{split}
\end{equation*}
The constant $\zeta(\mathbf{f})$ in \eqref{eq:resolve} is
uniquely determined by the orthogonality condition $\mathbf{u}\perp \mathbf{\Psi}_1$.
It follows from the bounds above that
$\zeta(\mathbf{f})$ is continuous linear functional on $L^2$.
\qed% end of the proof

Now we give a proof of Lemma~\ref{lem:l1-est}.

\noindent\textbf{Proof of Lemma~\ref{lem:l1-est}.}
Since $Y$ is continuously embedded into $L^2$,
the solution $\mathbf{u}=L^{-1}\mathbf{f}$ can be written as \eqref{eq:resolve} and
$$
\left\|\zeta(\mathbf{f})\mathbf{\Psi}_1\right\|_X \lesssim \|\mathbf{f}\|_{L^2}
\lesssim \|\mathbf{f}\|_{Y}.
$$

Next we estimate the second term of \eqref{eq:resolve}.
Noting that  $\|a\mathbf{\Psi}_1\|_X\le 2\|a\|_{L^\infty}$ for any $a\in L^\infty(\R)$,
we have
\begin{align*}
& \left\|\mathbf{\Psi}_1(x) \int_x^\infty e^{y/2}(e^{-2y}-2y)\sech(y) f_1(y)dy
\right\|_X \lesssim
 \left\|\int_x^\infty e^{y/2}(e^{-2y}-2y)\sech(y) f_1(y)dy\right\|_{L^\infty}
\\ & \le \inf_{f_1=g_1+h_1} \bigl(
\|g_1e^{-y/2}\cosh(y) \|_{L^2}\|\sech^2(y) (e^{-y}-2ye^y)\|_{L^2} 
\\ & +
 \|h_1e^{y/2}\cosh(y) \|_{L^1}\|\sech^2(y) (e^{-2y}-2y)\|_{L^\infty}\bigr)
\\ & \lesssim
 \inf_{f_1=g_1+h_1}\left(\|g_1e^{-y/2}\cosh(y)\|_{L^2}+\|h_1e^{y/2}\cosh(y)\|_{L^1}\right)
\lesssim \|f_1\|_{Y_1}\,.
\end{align*}

Similarly, we have
\begin{equation*}
\left\|\mathbf{\Psi}_1(x)
\int_{-\infty}^xe^{-y/2}(e^{2y}+2+2y)\sech(y) f_2(y)dy \right\|_X \lesssim \|f_2\|_{Y_2}.
\end{equation*}

Finally, we will estimate the fourth term of \eqref{eq:resolve}.
Clearly,
$$
\left\|\mathbf{\Psi}_2(x)\int^{\pm\infty}_x \mathbf{f}(y)\cdot\mathbf{\Theta}(y)dy
\right\|_{X} \le II_1+II_2+II_3+II_4\,,$$
where
\begin{align*}
II_{1}= & \left\|e^x\int^\infty_x \mathbf{f}(y)\cdot\mathbf{\Theta}(y) dy
\right\|_{L^2\cap L^\infty}\,,\\
II_{2}=& 2\left\|(1+x)\int^x_{\pm\infty}
\mathbf{f}(y)\cdot\mathbf{\Theta}(y) dy\right\|_{L^\infty}\,, \\
II_{3}=& \left\|e^{-x} \int_{-\infty}^x \mathbf{f}(y)\cdot\mathbf{\Theta}(y) dy
\right\|_{L^2\cap L^\infty}\,,\\
II_{4}=& 2\left\|x \int_{\pm\infty}^x
\mathbf{f}(y)\cdot\mathbf{\Theta}(y) dy\right\|_{L^\infty}\,.
\end{align*}
Since $\|e^{|y|/2}\mathbf{f}\|_{L^2}\lesssim \|\mathbf{f}\|_Y$
and $|\mathbf{\Theta}(y)|\lesssim e^{-|y|/2}$ for all $y \in \R$, we have
\begin{align*}
II_1 \le &\left\|\int^\infty_x e^{x-y} e^{|y|/2}(|f_1(y)|+|f_2(y)|)dy
\right\|_{L^2\cap L^\infty}
\\ \le & (\|e^{|y|/2}f_1\|_{L^2}+\|e^{|y|/2}f_2\|_{L^2})\|e^{-x}\|_{L^1(\R_+)\cap L^2(\R_+)}
\lesssim \|\mathbf{f}\|_{Y}\,.
\end{align*}
Similarly, we have $II_2+II_3+II_4\lesssim \|\mathbf{f}\|_{Y}$. Therefore
$$
\left\|\mathbf{\Psi}_2(x)\int^{\pm\infty}_x \mathbf{f}(y)\cdot\mathbf{\Theta}(y)dy
\right\|_{X} \lesssim \|\mathbf{f}\|_Y.$$
Thus we prove Lemma~\ref{lem:l1-est}.
\qed% end of the proof

Now we are in position to prove Lemma~\ref{lem:growth-rate}.

\noindent\textbf{Proof of Lemma~\ref{lem:growth-rate}.}
Let $\mathbf{\Psi}$ be a solution of the system \eqref{Lax-1-converse} in Lemma~\ref{lemma-1-converse}
such that
$$
\mathbf{\Psi}=\mathbf{\Psi}_1+\mathbf{\Phi}\,,\quad
\langle \mathbf{\Phi}\,,\, \mathbf{\Psi}_1 \rangle_{L^2}=0\,.
$$
Substituting \eqref{eq:Qdecomp1} (with $k=1$ and $v = \gamma=\theta=0$) into
the system \eqref{Lax-1-converse}, we obtain
\begin{equation}
  \label{eq:Lax-1-c2}
L\mathbf{\Phi} =\mathbf{R}_1+\mathbf{R}_2+R_3\mathbf{\Phi}\,,
\end{equation}
where
\begin{align*}
\mathbf{R}_1 & =  M(S)\mathbf{\Psi}_1
= \begin{bmatrix}SQ_1e^{x/2} \\ -\overline{S}Q_1e^{-x/2}\end{bmatrix}\,, \\
\mathbf{R}_2 & =  \left[ -\alpha xQ_1\sigma_2+\beta(x \partial_x Q_1+Q_1)\sigma_1 \right]
\mathbf{\Psi}_1 = i\alpha xQ_1^2\begin{bmatrix}e^{x/2}\\ e^{-x/2}\end{bmatrix}
+\beta(x \partial_x Q_1 +Q_1)Q_1\begin{bmatrix}e^{x/2}\\ -e^{-x/2}\end{bmatrix} \,, \\
R_3 & = M(S)-\alpha xQ_1\sigma_2+\beta(x \partial_x Q_1 +Q_1)\sigma_1\,.
\end{align*}

Because $\mathbf{\Psi}_1 \notin Y$ and $\|(I-P) {\bf f}\|_{Y}=\infty$ whatever
${\bf f}$ is, we shall modify the projection operator compared to the proof of Lemma~\ref{lemma-1-converse}.
Let $\widetilde{P}:L^2(\R;\C^2)\to L^2(\R;\C^2) \cap {}^\perp\ker(L^*)$ be a new projection
defined by
$$
\widetilde{P}\mathbf{u}=\mathbf{u}-\frac{3}{4}
\langle\mathbf{u},\mathbf{\Theta}\rangle_{L^2} \sech^2(x)\,\mathbf{\Psi}_1\,.
$$

Since $\Re \langle S,Q_1\rangle_{L^2}=\Im \langle S, \partial_x Q_1\rangle_{L^2}=0$ by
\eqref{eq:Qdecomp2}, we have
\begin{equation}
\label{eq:cancel2}
\left\langle M(S)\mathbf{\Psi}_1\,,\mathbf{\Theta}\right\rangle_{L^2}=
-2\Re\langle S,Q_1\rangle_{L^2}+2i\Im\langle S, \partial_x Q_1\rangle_{L^2} =0\,.
\end{equation}
By \eqref{eq:cancel2} and the fact that
$\mathbf{\Theta}\perp\operatorname{Range}(L)$, we obtain
\begin{align*}
L\mathbf{\Phi} = \widetilde{P}L\mathbf{\Phi}
= \mathbf{R}_1+\widetilde{P}(\mathbf{R}_2+R_3\mathbf{\Phi})\,.
\end{align*}
Thus, the system \eqref{eq:Lax-1-c2} is transformed into
\begin{equation}
  \label{eq:Lax-1-c3}
(I-L^{-1}\widetilde{P}R_3)\mathbf{\Phi}
=L^{-1}\mathbf{R}_1+L^{-1}\widetilde{P}\mathbf{R}_2\,.
\end{equation}
Lemma~\ref{lem:l1-est} and the bound \eqref{eq:Qdecomp3} imply
$$  \|L^{-1}\mathbf{R}_1\|_X\lesssim \|\mathbf{R}_1\|_{Y}
 \lesssim  \|SQ_1\cosh(x) \|_{L^2}\lesssim \|S\|_{L^2}\,,$$
\begin{align*}
\|L^{-1}\widetilde{P}\mathbf{R}_2\|_X\lesssim &
 \|\widetilde{P}\mathbf{R}_2\|_{Y}
\lesssim
\|\mathbf{R}_2\|_{Y}+\left|\langle \mathbf{R}_2\,,\,\mathbf{\Theta}\rangle_{L^2}\right|
\\ \lesssim & |\alpha|\|x Q_1^2\cosh(x)\|_{L^2}
+|\beta|\|(x \partial_x Q_1 +Q_1)Q_1\cosh(x)\|_{L^2}+\|\mathbf{R}_2\|_{L^2} \\
\lesssim & \|Q-Q_1\|_{L^2},
\end{align*}
and for $\mathbf{u}\in X$,
\begin{align*}
& \|L^{-1}\widetilde{P}R_3\mathbf{u}\|_{X}
\lesssim  \|\widetilde{P}R_3\mathbf{u}\|_{Y}\\
\lesssim & 
\|(\Re S)\sigma_1\mathbf{u}\|_Y+\|(\Im S)\sigma_2\mathbf{u}\|_Y
+|\alpha|\|xQ_1\sigma_2\mathbf{u}\|_Y+|\beta|\|(x \partial_x Q_1+Q_1)\sigma_1\mathbf{u}\|_Y
+\|R_3\mathbf{u}\|_{L^1}
\\ \lesssim & (\|S\|_{L^2}+|\alpha|\|xQ_1\|_{L^2}+|\beta|\|x \partial_x Q_1 +Q_1\|_{L^2})
\|\mathbf{u}\|_X \\
\lesssim & \|Q-Q_1\|_{L^2}\|\mathbf{u}\|_X \,.
\end{align*}
If $\|Q-Q_1\|_{L^2}$ is sufficiently small, then
$I-L^{-1}\widetilde{P}R_3$ is invertible on $X$ and
$$
\|\mathbf{\Phi} \|_{X}\le  \|(I-L^{-1}\widetilde{P}R_3)^{-1}
(L^{-1}\mathbf{R}_1+L^{-1}\widetilde{P}\mathbf{R}_2)\|_{X}
\lesssim  \|Q-Q_1\|_{L^2}\,.$$
Thus we prove \eqref{eq:Psidecomp2}.

Next, we will prove \eqref{eq:Psidecomp2'}.
Differentiating \eqref{eq:Lax-1-c2} $m$ times $(0\le m\le n)$, we have
\begin{equation}
\label{eq:***}
L\partial_x^m \mathbf{\Phi} = \partial_x^m(\mathbf{R}_1+\mathbf{R}_2+R_3 \mathbf{\Phi})
+\left[L\,,\, \partial_x^m\right] \mathbf{\Phi}\,.
\end{equation}

Let $\hat{P} : L^2(\R;\C^2)\to L^2(\R;\C^2)\cap {}^\perp\ker(L)$ be another
projection defined by
$$
\hat{P} \mathbf{u}=\mathbf{u}-\frac{1}{\sqrt{2\pi}}
\langle \mathbf{u},\mathbf{\Psi}_1\rangle_{L^2} \mathbf{\Psi}_1\,.
$$
where we used $\| \mathbf{\Psi}_1 \|_{L^2}^2 = 4 \int_0^{\infty} \sech(x) dx=2\pi$.
Since $L = L \hat{P} = \widetilde{P} L \hat{P}$, equation
\eqref{eq:***} can be rewritten as
\begin{gather*}
(L-\widetilde{P}R_3) \hat{P} \partial_x^m \mathbf{\Phi}
=\widetilde{P}\mathbf{R}_{4,m}\,,
\end{gather*}
where
$\mathbf{R}_{4,m}= \partial_x^m\left(\mathbf{R}_1+\mathbf{R}_2\right)
+\left\{[\partial_x^m,Q_1\sigma_1]+[\partial_x^m,R_3]
+R_3[\partial_x^m, \hat{P}]\right\} \mathbf{\Phi}$.
Note that $\hat{P}\mathbf{\Phi}=\mathbf{\Phi}$.
Suppose that $\|\partial_x^l \mathbf{\Phi} \|_X\lesssim \|Q-Q_1\|_{H^l}
+\|Q-Q_1\|_{H^l}^l$ for $0\le l< m \le n$.
Then by the induction hypothesis, we have
$$
\|\mathbf{R}_{4,m}\|_Y\lesssim \|Q-Q_1\|_{H^m}+\|Q-Q_1\|_{H^m}^m.
$$
Therefore, if $\|Q-Q_1\|_{L^2}$ is sufficiently small, then
$I-L^{-1}\widetilde{P}R_3$ is invertible on $X$ and
\begin{align*}
\|\hat{P} \partial_x^m \mathbf{\Phi} \|_{X}\le &
\|(I-L^{-1}\widetilde{P}R_3)^{-1}L^{-1}\mathbf{R}_{4,m}\|_{X}
\lesssim \|\mathbf{R}_{4,m}\|_Y\lesssim \|Q-Q_1\|_{H^m}
+\|Q-Q_1\|_{H^m}^m\,,
\end{align*}
and
\begin{align*}
 \|\partial_x^m \mathbf{\Phi}\|_{X} \le &
\|\hat{P} \partial_x^m \mathbf{\Phi}\|_{X}+\|[\partial_x^m,\hat{P}] \mathbf{\Phi}\|_{X}
\\ \lesssim & \|\hat{P} \partial_x^m \mathbf{\Phi}\|_{X}+\|\mathbf{\Phi}\|_{L^2}
\\ \lesssim & \|Q-Q_1\|_{H^m}+\|Q-Q_1\|_{H^m}^m\,.
\end{align*}
This  completes the proof of Lemma~\ref{lem:growth-rate}.
\qed% end of the proof

\section{From the zero solution to a $1$-soliton}
\label{sec:backlund-time-evolution}

In this section, we will prove Theorem~\ref{theorem-main} by showing
that a B\"acklund transformation (\ref{new-solution}) maps smooth
solutions of \eqref{NLS-integrable} in an $L^2$-neighborhood of the zero solution to
those in an $L^2$-neighborhood of a $1$-soliton.

First of all, we construct a fundamental system of solutions of the spectral problem
 (\ref{Lax-1}) with $\eta=\frac12$, which will be assumed throughout this section.
If $q=0$, the fundamental system of solutions of (\ref{Lax-1}) with $\eta=\frac12$
is given by the two solutions
\begin{equation}
\label{eq:fsystemq0}
\mbox{\boldmath $\psi$}_1(x) = \begin{bmatrix}  e^{x/2} \\ 0\end{bmatrix}\,, \quad
\mbox{\boldmath $\psi$}_2(x) = \begin{bmatrix}  0 \\ -e^{-x/2} \end{bmatrix}.
\end{equation}
When $q$ is small in $L^2$, a fundamental system of \eqref{Lax-1} with $\eta=\frac12$
can be found as a perturbation of the two linearly independent solutions \eqref{eq:fsystemq0}.

Let us consider the following boundary value problems
\begin{equation}
  \label{eq:bc1}
\left\{
  \begin{aligned}
& \varphi_1'=q\varphi_2\,,\\
& \varphi_2'=-\bar{q}\varphi_1-\varphi_2\,,\\
& \lim_{x\to\infty}\varphi_1(x)=1\,, \\ 
& \lim_{x\to-\infty}e^{x}\varphi_2(x)=0,
  \end{aligned}\right.
\end{equation}
and
\begin{equation}
  \label{eq:bc2}
\left\{
  \begin{aligned}
& \chi_1'=\chi_1+q\chi_2\,,\\
& \chi_2'=-\bar{q}\chi_1\,,\\
& \lim_{x\to\infty}e^{-x}\chi_1(x)=0\,, \\
& \lim_{x\to-\infty}\chi_2(x)=-1\,.
  \end{aligned}\right.
\end{equation}
If the boundary value problems \eqref{eq:bc1} and \eqref{eq:bc2} have a unique
solution, then
\begin{equation}
\label{eq:fsystemq1}
\bm{\psi}_1(x) = e^{x/2} \begin{bmatrix}\varphi_1(x)\\ \varphi_2(x)\end{bmatrix}, \quad
\bm{\psi}_2(x) = e^{-x/2}\begin{bmatrix}\chi_1(x)\\ \chi_2(x)\end{bmatrix}
\end{equation}
become linearly independent solutions of the system \eqref{Lax-1} with $\eta=\frac12$.
It follows from a standard ODE theory that
every solution of the system \eqref{Lax-1} with $q\in C(\R)$ can be written as
a linear superposition of the two solutions (\ref{eq:fsystemq1}).

Uniqueness of solutions of the boundary value problems \eqref{eq:bc1} and \eqref{eq:bc2}
follows from the following lemma.

\begin{lemma}
\label{lem:Lax1,0}
There exists a $\delta>0$ such that if $\|q\|_{L^2}<\delta$, then
the boundary value problems \eqref{eq:bc1} and \eqref{eq:bc2} have a solution
in the class
$$(\varphi_1\,,\varphi_2)\in L^\infty\times(L^2\cap L^\infty)\,,
\quad (\chi_1\,,\chi_2)\in (L^2\cap L^\infty)\times L^\infty.$$
Moreover, there exists a $C>0$ such that
\begin{gather*}
\|\varphi_1-1\|_{L^\infty}+\|\varphi_2\|_{L^2\cap L^\infty}\le C\|q\|_{L^2},\\
\|\chi_1\|_{L^2\cap L^\infty}+\|\chi_2+1\|_{L^\infty}\le C\|q\|_{L^2}.
\end{gather*}
\end{lemma}

\begin{proof}
Let us translate the boundary value problem \eqref{eq:bc1} into a system of integral equations
\begin{equation}
\label{eq:intphi1}
\left\{ \begin{array}{l}
\varphi_1(x)=1-\int_x^\infty q(y)\varphi_2(y)dy=:T_1(\varphi_1,\varphi_2)(x),\\
\varphi_2(x)=-\int_{-\infty}^xe^{-(x-y)}\overline{q(y)}\varphi_1(y)dy
=:T_2(\varphi_1,\varphi_2)(x). \end{array} \right.
\end{equation}
Let us introduce a Banach space $Z := L^\infty\times(L^\infty\cap L^2)$ equipped
with the norm
\begin{gather*}
\|(u_1,u_2)\|_Z=\|u_1\|_{ L^\infty}+\|u_2\|_{L^\infty\cap L^2}\,.
\end{gather*}
In order to find a solution of the system \eqref{eq:intphi1},
we will show that $T=(T_1,T_2):Z\to Z$ is a contraction mapping.

Using the Schwarz inequality and Young's inequality, we have for $(\varphi_1,\varphi_2)$ and $(\tilde{\varphi}_1,\tilde{\varphi}_2) \in Z$,
\begin{align*}
\|T_1(\varphi_1,\varphi_2)-T_1(\tilde{\varphi}_1,\tilde{\varphi}_2)\|_{L^\infty}
= \sup_{x \in \R} \left|\int_x^\infty q(y)(\varphi_2(y)-\tilde{\varphi}_2(y))dy\right|
\le \|q\|_{L^2}\|\varphi_2-\tilde{\varphi}_2\|_{L^2}\,,
\end{align*}
and
\begin{align*}
\|T_2(\varphi_1,\varphi_2)- T_2(\tilde{\varphi}_1,\tilde{\varphi}_2)\|_{L^2\cap L^\infty}
= & \left\|\int^x_{-\infty}e^{-(x-y)} \overline{q(y)}(\varphi_1(y)-\tilde{\varphi_1}(y))dy\right\|_{L^2\cap L^\infty}
\\ \le & \|e^{-x}\|_{L^1(\R_+) \cap L^2(\R_+)}\|q\|_{L^2}\|\varphi_1-\tilde{\varphi_1}\|_{L^\infty}
\\ \le & \|q\|_{L^2}\|\varphi_1-\tilde{\varphi_1}\|_{L^\infty}
\end{align*}
If $\|q\|_{L^2}$ is sufficiently small, then $T=(T_1,T_2)$ is a contraction
mapping on $Z$. Therefore $T=(T_1\,,T_2)$ has a unique  fixed point
$(\varphi_1,\varphi_2) \in Z$ and
\begin{align*}
\|\varphi_1-1\|_{L^\infty}+\|\varphi_2\|_{L^\infty\cap L^2}
= & \|T(\varphi_1,\varphi_2)-T(0,0)\|_Z
\le \|q\|_{L^2}\|(\varphi_1,\varphi_2)\|_Z
\\ \le & \|q\|_{L^2}(1+\|\varphi_1-1\|_{L^\infty}+\|\varphi_2\|_{L^\infty\cap L^2})\,.
\end{align*}
Thus we have
$$
\|\varphi_1-1\|_{L^\infty}+\|\varphi_2\|_{L^\infty}+\|\varphi_2\|_{L^2} = {\cal O}(\|q\|_{L^2}).
$$
Finally, we confirm the boundary conditions in the system \eqref{eq:bc1}.
By \eqref{eq:intphi1} and the fact that $q\in L^2$ and $\varphi_2\in L^2$,
we have $\lim_{x\to\infty}\varphi_1(x)=1$.
Since $\varphi_2$ is bounded and continuous, it is clear that
$\lim_{x\to-\infty}e^x\varphi_2(x)=0$.

In the same way, we can prove that the boundary value problem \eqref{eq:bc2} has a unique solution
$(\chi_1,\chi_2) \in \widetilde{Z} := (L^\infty\cap L^2)\times L^\infty $ satisfying
$$
\|\chi_1\|_{L^2\cap L^\infty}+\|\chi_2+1\|_{L^\infty} = {\cal O}(\|q\|_{L^2})
$$
and the boundary conditions $\lim_{x\to\infty}e^{-x}\chi_1(x)=0$ and
$\lim_{x\to-\infty}\chi_2(x)=-1$.
\end{proof}

Next we will consider the time evolution of $(\psi_1,\psi_2)$.
We will evolve $(\psi_1,\psi_2)$ by the linear time evolution \eqref{Lax-2}
for initial data $(\psi_1(0,x),\psi_2(0,x))$ satisfying the spectral problem
\eqref{Lax-1} at $t=0$ assuming that $q(t,x)$ is a solution of
\eqref{NLS-integrable}.

Suppose  that $\bm{\varphi}(t,x)={}^t(\varphi_1(t,x),\varphi_2(t,x))$ satisfies
the boundary value problem \eqref{eq:bc1} at $t=0$ with $q = q(0,x)$
and that $e^{x/2}\bm{\varphi}(t,x)$ satisfies \eqref{Lax-2} for every $t\ge0$
and $x\in\R$. Then the  linear time evolution of $\bm{\varphi}(t,x)$ can be
written in the matrix form
\begin{equation}
\label{eq:phi-st1}
\partial_t \bm{\varphi}(t,x) = A(t,x) \bm{\varphi}(t,x)\,, \quad
A(t,x)=\begin{bmatrix}a(t,x) & b(t,x) \\ c(t,x) & -a(t,x) \end{bmatrix}\,,
\end{equation}
where
$$
a = i \left(\frac12+|q|^2\right),\quad
b=i( \partial_x q + q),\quad c = i(\partial_x \bar{q} - \bar{q})\,.
$$
Similarly, let $\bm{\chi}(t,x)={}^t(\chi_1(t,x),\chi_2(t,x))$ be a solution of
the boundary value problem \eqref{eq:bc2} at $t=0$ with $q = q(0,x)$
whose time evolution is written in the same matrix form (\ref{eq:phi-st1}) for
$\bm{\chi}(t,x)$. Solutions $\bm{\varphi}(t,x)$ and $\bm{\chi}(t,x)$
are characterized by the following lemma.

\begin{lemma}
\label{lem:Lax1,t}
Suppose that $q\in C(\R;H^3(\R))$ is a solution of \eqref{NLS-integrable}
and that $\|q(0,\cdot)\|_{L^2}$ is sufficiently small.
Let $\bm{\varphi}={}^t(\varphi_1,\varphi_2)$ and $\bm{\chi}={}^t(\chi_1,\chi_2)$
be solutions of the linear equation
\eqref{eq:phi-st1} starting with the initial data given by solutions of
the boundary value problems \eqref{eq:bc1} and \eqref{eq:bc2} respectively
with $q = q(0,x)$. Then $\partial_x^i\bm{\varphi}\in C(\R; Z)$ and
$\partial_x^i\bm{\chi}\in C(\R;\widetilde{Z})$ for $0\le i\le 3$ and
for every $t\in\R$,
\begin{equation}
\label{eq:phi-st2}
\left\{   \begin{aligned}
& \partial_x\varphi_1(t,x)=q(t,x)\varphi_2(t,x)\,,\\
& \partial_x\varphi_2(t,x)=-\overline{q(t,x)}\varphi_1(t,x)-\varphi_2(t,x)\,, \\ 
& \lim_{x\to\infty}\varphi_1(t,x)=e^{it/2}\,, \\
& \lim_{x\to-\infty}e^x\varphi_2(t,x)=0\,, \end{aligned} \right.
\end{equation}
and
\begin{equation}
\label{eq:chi-st2} \left\{   \begin{aligned}
& \partial_x\chi_1(t,x)=\chi_1(t,x)+q(t,x)\chi_2(t,x)\,,\\
& \partial_x\chi_2(t,x)=-\overline{q(t,x)}\chi_1(t,x)\,,\\
& \lim_{x\to\infty}e^{-x}\chi_1(t,x)=0\,, \\
& \lim_{x\to-\infty}\chi_2(t,x)=-e^{-it/2}\,.\end{aligned} \right.
\end{equation}
\end{lemma}

\begin{proof}
First, we will prove that the boundary value problem
\eqref{eq:phi-st2} holds for every $t \in \R$.

The coefficient matrix $A(t,x)$ of the system \eqref{eq:phi-st1} is
continuous in $(t,x)$ and $C^1$ in $x$ since $q(t,x)\in
C(\R;H^3(\R))$.  By a bootstrapping argument for the system
\eqref{eq:bc1}, Lemma~\ref{lem:Lax1,0} implies that $\varphi_1(0,x)$
and $\varphi_2(0,x)$ are $C^1$ in $x$.  Solving the Cauchy problem for
the linear evolution equation \eqref{eq:phi-st1}, we find that
$\varphi_1(t,x)$ and $\varphi_2(t,x)$ are in $C^1(\R \times \R)$.  By
a bootstrapping argument for the systems \eqref{eq:bc1} and
\eqref{eq:phi-st1}, we conclude that $\partial_x\partial_t\bm{\varphi}(t,x)$ and
$\partial_t\partial_x\bm{\varphi}(t,x)$ are in $C(\R \times \R;\R^2)$ and thus
$\partial_x\partial_t\bm{\varphi}(t,x) = \partial_t\partial_x\bm{\varphi}(t,x)$.

Let
 \begin{gather*}
B(t,x)=\begin{bmatrix}0 & q(t,x) \\ -\overline{q(t,x)} & -1\end{bmatrix}\,,
\quad
\mathbf{F}(t,x)=\partial_x\bm{\varphi}(t,x)-B(t,x)\bm{\varphi}(t,x)\,.
\end{gather*}
Since $q$ is a solution of \eqref{NLS-integrable},
the matrices $A$ and $B$ satisfy the Zakharov-Shabat compatibility condition
\begin{equation}
\label{eq:compatibility}
\partial_xA - \partial_tB + [A\,,B] = 0\,.
\end{equation}
As a result, we obtain
\begin{align*}
\partial_t\mathbf{F}=& \partial_t\partial_x\bm{\varphi}-(\partial_tB)\bm{\varphi}-B\partial_t\bm{\varphi} = \partial_x(A\bm{\varphi})-(\partial_tB)\bm{\varphi}-BA\bm{\varphi}
\\=& (\partial_xA+[A\,,\,B]-\partial_tB)\bm{\varphi}+A\mathbf{F} = A\mathbf{F}\,.
\end{align*}
Applying Gronwall's equality, we see that for any $T>0$,
there exists a constant $C(T)$ such that
$$
|\mathbf{F}(t)|\le C(T)|\mathbf{F}(0)|, \quad t\in[-T,T].
$$
Since $\mathbf{F}(0)=\mathbf{0}$ by the assumption,
it follows that $\mathbf{F}(t)=\mathbf{0}$
for every $t\in\R$. Thus we prove the differential part
of the system \eqref{eq:phi-st2}.

Next we will prove $\varphi_1(t,\cdot)\in L^\infty(\R)$ and
$\varphi_2(t,\cdot)\in L^2(\R) \cap L^\infty(\R)$ for every $t \in \R$.
By the linear evolution \eqref{eq:phi-st1}, we have
\begin{eqnarray*}
 \left| \partial_t(|\varphi_1(t,x)|^2+|\varphi_2(t,x)|^2)\right| & = &
4\left|\Im q(t,x)\overline{\varphi_1(t,x)}\varphi_2(t,x)\right| \\
& \le & 2\|q(t,\cdot) \|_{L^\infty}(|\varphi_1(t,x)|^2+|\varphi_2(t,x)|^2).
\end{eqnarray*}
Applying Gronwall's inequality again, we have
\begin{equation}
  \label{eq:gwphi}
|\varphi_1(t,x)|^2+|\varphi_2(t,x)|^2 \le
e^{\alpha |t|}(|\varphi_1(0,x)|^2+|\varphi_2(0,x)|^2), \quad t \in \R
\end{equation}
where $\alpha = 2\sup_{(t,x) \in \R \times \R} |q(t,x)|$. Since
$\bm{\varphi}(0,\cdot)\in L^\infty(\R;\C^2)$, bound \eqref{eq:gwphi} shows that
$\bm{\varphi}(t,\cdot)\in L^\infty(\R;\C^2)$ for any $t\in\R$.

Using the linear system \eqref{eq:phi-st1} again, we have
\begin{eqnarray*}
\partial_t|\varphi_2(t,x)|^2 & \le & 2| \partial_x q(t,x)-q(t,x)|\varphi_1(t,x)||\varphi_2(t,x)|
\\ & \le & |\varphi_2(t,x)|^2+|\varphi_1(t,x)|^2|\partial_x q(t,x)-q(t,x)|^2.
\end{eqnarray*}
By Gronwall's inequality, for any $T>0$ there exists a $C(T) >0$ such that
$$
|\varphi_2(t,x)|^2\le |\varphi_2(s,x)|^2 +C(T) \int_s^t
|\partial_x q(\tau,x)-q(\tau,x)|^2d\tau, \quad 0\le s\le t\le T, \quad x \in \R.
$$
Therefore, we have
\begin{equation}
\label{eq:gwphi2}
\|\varphi_2(t,\cdot)\|_{L^2}^2\le \|\varphi_2(s,\cdot)\|_{L^2}^2
+C(T)\int_s^t\|\partial_x q(\tau,\cdot)-q(\tau,\cdot)\|_{L^2}^2d\tau.
\end{equation}
Since  $\varphi_2(0,\cdot)\in L^2(\R)$, bound
\eqref{eq:gwphi2} shows that  $\varphi_2(t,\cdot)\in L^2(\R)$ for every $t\in\R$ and
$\|\varphi_2(t)\|_{L^2}$ is continuous in $t$.
Since $A(t,\cdot)\in C(\R;H^2(\R))$ and $\|\varphi_1(t)\|_{L^\infty}$ and
$\|\varphi_2(t)\|_{L^2\cap L^\infty}$ are bounded locally in time,
the linear system \eqref{eq:phi-st1} implies that
$\varphi_1(t,\cdot)$  and $\varphi_2(t,\cdot)$ are continuous in $L^\infty(\R)$
and thus $\varphi_2(t,\cdot)$ is continuous in $L^2(\R)$.
Using the fact that $\bm{\varphi}\in C(\R;Z)$ and a bootstrapping argument
for the system \eqref{eq:bc1}, we have $\partial_x^i\bm{\varphi} \in C(\R;Z)$ for $1\le i\le 3$.

It remains to prove the boundary conditions of the system \eqref{eq:phi-st2}.
Since $\varphi_2(t,x)$ is bounded and continuous in $x$ for every fixed $t\in\R$,
we have $\lim_{x\to-\infty}e^x\varphi_2(t,x)=0$.
By a variation of constants formula, we have
\begin{equation}
  \label{eq:varfphi}
\bm{\varphi}(t,x)=e^{i\sigma_3t/2}\bm{\varphi}(0,x)
+\int_0^t e^{i\sigma_3(t-s)/2}A_1(s,x)\bm{\varphi}(s,x)ds, \quad
\end{equation}
where $A_1(t,x)=A(t,x)-i\sigma_3/2$.
By the assumption that $q\in C(\R;H^3(\R))$, we have
$$
\sup_{x\in\R}\sup_{0\le s\le t}|A_1(s,x)|<\infty \quad \mbox{\rm and} \quad
\lim_{x\to\pm\infty}A_1(s,x)=0.
$$
Applying  Lebesgue's dominated convergence theorem to
the integral equation \eqref{eq:varfphi}, we get
$$
\lim_{x\to\infty}\left|\bm{\varphi}(t,x)-e^{i\sigma_3t/2}\bm{\varphi}(0,x)\right| = 0\,.
$$
Combining the above with the boundary condition $\lim_{x\to\infty}\varphi_1(0,x)=1$,
we obtain 
$$
\lim_{x\to\infty}\varphi_1(t,x)=e^{it/2}.
$$

Properties of $\bm{\chi}$ and the boundary value problem \eqref{eq:chi-st2}
can be proven in the same way as properties of $\bm{\varphi}$
and the boundary value problem \eqref{eq:phi-st2}.
\end{proof}

Now, we have time global estimates of solutions to the linear evolution equation \eqref{eq:phi-st1}.

\begin{lemma}
\label{lem:phi-bdt}
Let $q\in C(\R;H^3(\R))$ be a solution of \eqref{NLS-integrable}.
Suppose that $\bm{\varphi}(t,x)$ and $\bm{\chi}(t,x)$ are solutions of the linear evolution equation
\eqref{eq:phi-st1} such that $\bm{\varphi}(0,x)\in Z$
and $\bm{\chi}(0,x)\in \widetilde{Z}$, respectively.
There exist  positive constants $\varepsilon$ and $C$ such that if
$\|q(0,\cdot)\|_{L^2}<\varepsilon$, then for every $t\in\R$,
\begin{gather}
  \label{eq:phi-bdt}
\|\varphi_1(t,\cdot)-e^{it/2}\|_{L^\infty}+\|\varphi_2(t,\cdot)\|_{L^2\cap L^\infty}
\le C\|q(0,\cdot)\|_{L^2},\\
  \label{eq:chi-bdt}
\|\chi_1(t,\cdot)\|_{L^2\cap L^\infty}+\|\chi_2(t,\cdot)+e^{-it/2}\|_{L^\infty}
\le C\|q(0,\cdot)\|_{L^2}.
\end{gather}
\end{lemma}

\begin{proof}
Since $\bm{\varphi}(t,\cdot)\in Z$ and $\bm{\chi}(t,\cdot)\in \widetilde{Z}$
for each $t \in \R$ and satisfy the boundary value problem \eqref{eq:phi-st2} and
\eqref{eq:chi-st2},
Lemma~\ref{lem:phi-bdt} can be proven in exactly the same way as
Lemma~\ref{lem:Lax1,0}.
\end{proof}

Our next result shows that the B\"{a}cklund transformation (\ref{new-solution})
with $\eta=\frac{1}{2}$ generates a new solution $Q$ in a $L^2$-neighborhood of the $1$-soliton
$e^{i(t+\theta)} Q_1(x-\gamma)$, where $Q_1(x) = \sech(x)$.

\begin{lemma}
\label{lem:profile}
Let $\varepsilon$ be a sufficiently small positive number.
Let $q(t,x)\in C(\R;H^3(\R))$ be a solution of \eqref{NLS-integrable} such that
$\|q(0,\cdot)\|_{L^2}<\varepsilon$ and let
\begin{eqnarray}
\label{eq:defpsitx}
\left\{ \begin{array}{l}
\psi_1(t,x)=c_1e^{x/2}\varphi_1(t,x)+c_2e^{-x/2}\chi_1(t,x)\,,\\
\psi_2(t,x)=c_1e^{x/2}\varphi_2(t,x)+c_2e^{-x/2}\chi_2(t,x)\,, \end{array} \right.
\end{eqnarray}
where $c_1=ae^{(\gamma+i\theta)/2}$, $c_2=ae^{-(\gamma+i\theta)/2}$ and $a\ne0$, $\gamma\in\R$, $\theta\in\R$
are constants. Let
\begin{equation}
\label{eq:defQtx}
Q(t,x)=-q(t,x)-\frac{2\psi_1(t,x)\overline{\psi_2(t,x)}}{|\psi_1(t,x)|^2+|\psi_2(t,x)|^2},
\end{equation}
Then $Q\in C(\R;H^3(\R))$ and $Q(t,x)$ is a solution of \eqref{NLS-integrable}.
Moreover, there is an $\varepsilon$-dependent constant $C > 0$ such that
\begin{equation}
  \label{eq:profile}
 \sup_{t\in\R}\|Q(t,\cdot) - e^{i(t+\theta)}Q_1(\cdot-\gamma)\|_{L^2} \leq C \|q(0,\cdot)\|_{L^2}\,.
\end{equation}
\end{lemma}

\begin{proof}
Since $\mbox{\boldmath $\psi$}$ in (\ref{eq:defpsitx}) solve the Lax system (\ref{Lax-1})
and (\ref{Lax-2}), the B\"{a}cklund transformation (\ref{eq:defQtx}) implies
that if $q(t,x)$ is a solution of \eqref{NLS-integrable}, so is $Q(t,x)$.
Let us still give a rigorous proof of this fact for the sake of self-containedness.
Let
$$
\Psi_1(t,x) := \frac{\overline{\psi_2(t,x)}}
{|\psi_1(t,x)|^2+|\psi_2(t,x)|^2},\quad
\Psi_2(t,x) := \frac{\overline{\psi_1(t,x)}}
{|\psi_1(t,x)|^2+|\psi_2(t,x)|^2}.
$$
Thanks to \eqref{eq:phi-bdt} and \eqref{eq:chi-bdt},
$\bm\psi\ne\mathbf{0}$ for any $(t,x)\in\R^2$, hence
$Q$ and  $\bm\Psi$ are well defined for every $t \in \R$.
Since $\partial_x^i\bm{\varphi}\in C(\R;Z)$  and
$\partial_x^i\bm{\chi}\in C(\R;\widetilde{Z})$ for any $0\le i\le 3$
and
$$
q\in C(\R;H^3(\R))\cap C^1(\R;H^1(\R)),
$$
it follows from the linear evolution equation \eqref{eq:phi-st1}
that $\bm\Psi$ is of the class $C^1$
and $\partial_x\partial_t \bm\Psi$ and $\partial_t\partial_x \bm\Psi$ are continuous. Moreover $Q(t,\cdot)\in C(\R;H^3(\R))$.

By a straightforward but lengthy computation, we show that
\begin{gather}
\label{eq:Phix}
\partial_x \begin{bmatrix} \Psi_1\\ \Psi_2\end{bmatrix}
=\begin{bmatrix}\frac{1}{2} & Q\\ -\bar{Q} &-\frac12\end{bmatrix}
\begin{bmatrix} \Psi_1\\ \Psi_2\end{bmatrix}\,,\\
\label{eq:Phit}
\partial_t\begin{bmatrix}
\Psi_1\\ \Psi_2\end{bmatrix}
=i\begin{bmatrix}\frac{1}{2} +|Q|^2 & \partial_x Q+ Q \\ \partial_x \bar{Q}-\bar{Q} &-\frac12-|Q|^2
\end{bmatrix} \begin{bmatrix} \Psi_1\\ \Psi_2\end{bmatrix}\,.
\end{gather}
It is clear that $\bm\Psi(x,t) \ne {\bf 0}$ for every $(t,x)\in \R \times \R$.
Combining \eqref{eq:Phix}, \eqref{eq:Phit} and the compatibility condition
$\partial_t\partial_x \bm\Psi = \partial_x\partial_t \bm\Psi$, we obtain $iQ_t+Q_{xx}+2|Q|^2Q=0$.

Now we will show the bound \eqref{eq:profile}. Let
\begin{equation}
\label{correction-R}
\begin{split}
R(t,x) :=& -Q(t,x)-q(t,x) \\=&
\frac{2 (c_1\varphi_1(t,x)+c_2e^{-x}\chi_1(t,x))
(\overline{c_1e^x\varphi_2(t,x)}+\overline{c_2\chi_2(t,x)})}
{|c_1\varphi_1(t,x) + c_2e^{-x} \chi_1(t,x)|^2 e^{x}
+ |c_1e^{x} \varphi_2(t,x) + c_2\chi_2(t,x)|^2 e^{-x}} = \frac{2 R_1}{R_2}\,,
\end{split}
\end{equation}
where
\begin{align*}
R_1 := &e^{x+\gamma}\varphi_1(t,x)\overline{\varphi_2(t,x)}
+e^{-x-\gamma}\chi_1(t,x)\overline{\chi_2(t,x)} + e^{i\theta}\varphi_1(t,x)\overline{\chi_2(t,x)}
+ e^{-i\theta}\chi_1(t,x)\overline{\varphi_2(t,x)}\,,\\
R_2 := &e^{x+\gamma}(|\varphi_1(t,x)|^2+|\varphi_2(t,x)|^2)
+ e^{-x-\gamma}(|\chi_2(t,x)|^2+|\chi_1(t,x)|^2)
\\ & +2\Re \left[e^{i\theta}(\varphi_1(t,x)\overline{\chi_1(t,x)}
+\varphi_2(t,x)\overline{\chi_2(t,x)})\right]\,.
\end{align*}

For $x\ge -\gamma$,
\begin{equation}
  \label{eq:R3}
R=\frac{2 e^{-x-\gamma+i\theta}\varphi_1(t,x)\overline{\chi_2(t,x)}}
{|\varphi_1(t,x)|^2+e^{-2(x+\gamma)}|\chi_2(t,x)|^2}
+ {\cal O}(|\varphi_2(t,x)|+e^{-x-\gamma}|\chi_1(t,x)|)
\end{equation}
since  $|\varphi_1|$, $|\chi_2|\sim1$ and $\varphi_2$, $\chi_1\sim0$
by Lemma~\ref{lem:phi-bdt}.
Similarly, for $x \leq -\gamma$,
\begin{equation}
\label{eq:R4}
R=\frac{2 e^{x+\gamma+i\theta}\varphi_1(t,x)\overline{\chi_2(t,x)}}
{|\chi_2(t,x)|^2+e^{2(x+\gamma)}|\varphi_1(t,x)|^2}
+ {\cal O}(|\chi_1(t,x)|+e^{x+\gamma}|\varphi_2(t,x)|)\,.
\end{equation}
Combining \eqref{eq:R3} and \eqref{eq:R4}, we get
\begin{align*}
& |R(t,x)+e^{i(t+\theta)}\sech(x+\gamma)| \\ \le &
 Ce^{-|x+\gamma|}(\|\varphi_1-e^{it/2}\|_{L^\infty}
+\|\chi_2+e^{-it/2}\|_{L^\infty})+C(|\varphi_2(t,x)|+|\chi_1(t,x)|)\,,
\end{align*}
where $C$ is a constant depending only on $\|q(0,\cdot)\|_{L^2}$.
Thus by Lemma~\ref{lem:phi-bdt}, there is $C > 0$ such that 
$$
\sup_{t\in\R}\|R(t,\cdot)+e^{i(t+\theta)}\sech(\cdot +\gamma)\|_{L^2} \leq 
C \|q(0,\cdot)\|_{L^2}\,.
$$
Combining the above with the $L^2$-conservation law
$\|q(t,\cdot)\|_{L^2}=\|q(0,\cdot)\|_{L^2}$ of \eqref{NLS-integrable},
we conclude that
\begin{align*}
\|Q(t,\cdot) - e^{i(t+\theta)} \sech(\cdot +\gamma)\|_{L^2}
\le \|R(t,\cdot) + e^{i(t+\theta)} \sech(\cdot+\gamma)\|_{L^2}+\|q(t,\cdot)\|_{L^2}
\lesssim \|q(0,\cdot)\|_{L^2}.
\end{align*}
This completes the proof of the bound \eqref{eq:profile}.
\end{proof}

\begin{remark}
To prove Lemmas~\ref{lem:Lax1,t} and \ref{lem:phi-bdt},
we require $H^3$-regularity of a solution $q$ to \eqref{NLS-integrable}.
The high regularity assumption on $q(t,x)$ is only used to
prove qualitative properties
on a solution $(\psi_1,\psi_2)$ of the Lax system (\ref{Lax-1})
and (\ref{Lax-2}) and has not been used quantitatively to prove the bounds 
\eqref{eq:phi-bdt} and \eqref{eq:chi-bdt}.
This is the reason why we can prove Theorem~\ref{theorem-main}
for any initial data satisfying $\|u(0,\cdot)-Q_1\|_{L^2}<\varepsilon$ 
by using an approximation argument.
\end{remark}

Now we are in position to prove Theorem~\ref{theorem-main}.

\noindent\textbf{Proof of Theorem~\ref{theorem-main}.}
Thanks to the scaling invariance of \eqref{NLS-integrable},
we may choose $k=1$, that is $Q_k=Q_1$.
\par
(Step 1): First, we will show \eqref{L2stab-result} assuming that 
$u(0,\cdot)\in H^3(\R)$.
Lemmas~\ref{lemma-1-converse} and  \ref{lemma-2-converse} imply that if
$Q=u(0,\cdot)\in H^3(\R)$ and $\|u(0,\cdot)-Q_1\|_{L^2}$ is sufficiently small,
then there exist a solution $\mathbf{\Psi}$ of the system
\eqref{Lax-1-converse} with $\eta=(k+iv)/2$ satisfying
$$
\exists C > 0 : \quad |k-1|+|v| \leq C \|u(0,\cdot)-Q_1\|_{L^2}\,.
$$
Letting
\begin{gather*}
  q_0(x)=-u(0,x)
-\frac{2k\Psi_1(x) \overline{\Psi_2(x)}}{|\Psi_1(x)|^2+|\Psi_2(x)|^2}\,,
\end{gather*}
and
\begin{gather*}
 \psi_{1,0}(x)=\frac{\overline{\Psi_2(x)}}{|\Psi_1(x)|^2 + |\Psi_2(x)|^2}\,,
\quad
\psi_{2,0}(x)=\frac{\Psi_1(x)}{|\Psi_1(x)|^2 + |\Psi_2(x)|^2}\,,
\end{gather*}
we see that $(\psi_{1,0},\psi_{2,0})$ is a solution of
the system \eqref{Lax-1} with $q=q_0$.
We may assume $k=1$ and $v=0$  without loss of generality thanks to
the change of variables in Remark~\ref{remark-change} 
and the invariance of \eqref{NLS-integrable} under the transformation
$$
\lambda\tilde{q}(\lambda^2(t+t_0),\lambda(x+x_0))=e^{i(vx/2-v^2t/4)}q(t,x-vt),$$
where $\lambda>0$ and $t_0$, $x_0$, $v\in\R$ are constants.

By the linear superposition principle, we can find complex constants $c_1$ and $c_2$
satisfying
$$\bm{\psi}_0={}^t(\psi_{1,0},\psi_{2,0})
=c_1e^{x/2}\bm{\varphi}(0,x)+c_2e^{-x/2}\bm{\chi}(0,x)\,.$$
Let $q(t,x)$ be a solution of \eqref{NLS-integrable} with
$q(0,x)=q_0(x)$ and let
$$
\bm{\psi}(t,x)={}^t(\psi_1(t,x),\psi_2(t,x))=c_1e^{x/2}\bm{\varphi}(t,x)
+c_2e^{-x/2}\bm{\chi}(t,x)\,.
$$
Lemma~\ref{lem:profile} implies that $\bm{\psi}(t,x)$
is a solution of the Lax system \eqref{Lax-1} and \eqref{Lax-2}
and that $Q(t,x)$ defined  by \eqref{eq:defQtx} satisfies the
 stability result \eqref{eq:profile}.
Since $Q(t,x)$ is a solution of \eqref{NLS-integrable}
in the class $C(\R;H^3(\R))$ and
\begin{align*}
Q(0,x)=& -q(0,x)-\frac{2\psi_1(0,x)\overline{\psi_2(0,x)}}
{|\psi_1(0,x)|^2+|\psi_2(0,x)|^2}
\\=&  -q_0(x)-\frac{2\Psi_1(x)\overline{\Psi_2(x)}}
{|\Psi_1(x)|^2+|\Psi_2(x)|^2} = u(0,x)
\end{align*}
by the definition, we have $Q(t,x)=u(t,x)$.
\par
(Step 2): Next, we prove \eqref{L2stab-result} for any $u(0,\cdot)$ which is 
sufficiently close to $Q_1$ in $L^2(\R)$.
Let $\delta_1=\|u(0,\cdot)-Q_1\|_{L^2}$. Let $u_{n,0}\in H^3(\R)$ ($n\in\N$) be a
sequence such that
\begin{equation*}
\lim_{n\to\infty}\|u_{n,0}-u(0,\cdot)\|_{L^2}=0\,,
\end{equation*}
and let $u_n(t,x)$ be a solution of \eqref{NLS-integrable} with
$u_n(0,x)=u_{0,n}(x)$.
In view of  the first step, we see there exist a positive constant $C$
and real numbers $k_n$, $v_n$, $t_n$, $x_n$ ($n\in \N$) such that
\begin{equation}
  \label{eq:unifb}
\sup_{t\in\R}\|u_n(t+t_n,\cdot+x_n) - Q_{k_n,v_n}\|_{L^2}
+ |k_n-1| + |v_n| + |t_n| + |x_n| \le C\|u_{0,n}-Q_1\|_{L^2}\,.
\end{equation}
By \eqref{eq:unifb}, there exist $k_0$, $v_0$, $t_0$, $x_0$ and
subsequences of $\{k_n\}$, $\{v_n\}$, $\{t_n\}$, $\{x_n\}$ such that
\begin{equation}
  \label{eq:conv}
\lim_{j\to\infty}k_{n_j}=k_0\,,\quad\lim_{j\to\infty}v_{n_j}=v_0\,,\quad
\lim_{j\to\infty}t_{n_j}=t_0\,,\quad\lim_{j\to\infty}x_{n_j}=x_0\,.  
\end{equation}
It follows from the main theorem in Tsutsumi \cite{T}
(see also Theorem 5.2 in \cite{LP}) that \eqref{NLS-integrable} is
$L^2$-well-posed in the class of solutions (\ref{class-solutions}).
Therefore combining \eqref{eq:unifb} and \eqref{eq:conv},
we obtain (\ref{L2stab-result}). Thus we complete the proof.
\qed% end of the proof

\section{Discussions}

We finish this article with three observations which are opened for further work.

{\bf 1.} The Cauchy problem associated with the generalized
nonlinear Schr\"{o}dinger equation (\ref{NLS-potential})
is well studied in the context of dispersive decay of small-norm solutions.
Since the decay rate of the $L^{\infty}-L^1$ norm for the semi-group
$$
S(t) := e^{-it (-\partial_x^2 + V(x))}, \quad t > 0
$$
is ${\cal O}(t^{-1/2})$, the nonlinear term $\| u(t,\cdot) \|^{2p}_{L^{\infty}}$ 
is absolute integrable if $p > 1$. The case $p = 1$
of the cubic NLS equation is critical with respect to this dispersive decay 
in the $L^{\infty}-L^1$ norm. The scattering theory for small solutions
in the supercritical case $p > 1$ was studied long ago \cite{CW,HT,GV,O}. The scattering theory was
extended to the critical ($p = 1$) and subcritical ($p = \frac{1}{2}$) cases
by Hayashi and Naumkin \cite{HN1,HN2} using more specialized properties of the fundamental
solutions generated by the semi-group $S(t)$.

In particular, Hayashi and Naumkin proved that if $q_0 \in H^1(\R)
\cap L^2_1(\R)$ and $\| q_0 \|_{H^1} + \| q_0 \|_{L^2_1} \leq
\varepsilon$ for sufficiently small $\varepsilon > 0$, then there
exists a unique global solution $q(t,\cdot) \in C(\mathbb{R};H^1(\R) \cap
L^2_1(\R))$ of (\ref{NLS-integrable}) with $q(0) = q_0$ such that
\begin{equation}
\label{small-decay}
\exists C > 0 : \quad \|q(t,\cdot)\|_{H^1} \leq C \varepsilon, \quad
\|q(t,\cdot)\|_{L^{\infty}} \leq C \varepsilon (1 + |t|)^{-1/2}, \quad t \in \mathbb{R}_+.
\end{equation}

Space $L^2_1(\R)$ is needed to control an initially small norm $\|q_0\|_{L^1}$.
Recall from inverse scattering (see, e.g.,
\cite{Ablowitz}) that if $\| q_0 \|_{L^1}$ is small, then the spectral
problem (\ref{Lax-1}) admits no isolated eigenvalue and produces no
soliton in $q(t,\cdot)$ as $t \to \infty$. In other words, $q(t,\cdot)$
contains
only the dispersive radiation part. Unfortunately, the norm $\| q(t,\cdot)
\|_{L^2_1}$ (and the norm $\| q(t,\cdot) \|_{L^1}$) may grow as $t \to
\infty$.  Indeed, it is shown in \cite{HN1} that there exists a small
$\varepsilon>0$ such that
$$\|(x+2it\partial_x)q(t,\cdot)\|_{L^2}\lesssim (1+|t|)^\varepsilon\,,$$
which implies that $\| q(t,\cdot) \|_{L^2_1} \geq C (1 + |t|)$
as $t \to \infty$ for some $C>0$.
\par

The possible growth of $\|q(t,\cdot)\|_{L^1}$ is an 
obstruction on the use of the B\"acklund transformation in  our approach.
If we can prove that the B\"{a}cklund transformation provides an
isomorphism between a ball $B_{\delta}(0) \ni q$ of small radius
$\delta > 0$ centered at $0$ in the energy space $H^1(\R)$ and a ball
$B_{\varepsilon}(Q_1) \ni Q$ of small radius $\varepsilon > 0$ centered at
$Q_1(x) = \sech(x)$ in the same energy space $H^1(\R)$ such that
$$
\exists C > 0: \quad \| Q - Q_1 \|_{L^{\infty}} \leq C \| q \|_{L^{\infty}},
$$
then the asymptotic stability of $1$-solitons holds in the following sense:
There exist positive constants $C$ and $\varepsilon$ such that if
$u(t,\cdot) \in C(\R_+,H^1(\R))$ is a solution of (\ref{NLS-integrable})
with $u(0) = u_0$ and $\| u_0 - Q_1 \|_{H^1 \cap L^2_1} \leq \varepsilon$, then
there exist constants $k \in \R$ and $v \in \R$ such that
\begin{equation}
\label{orbital-stab}
|k - 1| \leq C \varepsilon, \quad |v| \leq C \varepsilon, \quad
\inf_{(t_0,x_0) \in \R^2} \| u(t,\cdot) - Q_{k,v}(t - t_0,\cdot - x_0) \|_{H^1} \leq C
\|u_0-Q_1\|_{H^1\cap L^2_1}\,,
\end{equation}
and
\begin{equation}
\label{asymptotic-stab}
\lim_{t \to \infty} \| u(t,\cdot) - Q_{k,v}(t - t_0',\cdot - x_0') \|_{L^{\infty}} = 0,
\end{equation}
where $(t_0',x_0')$ are optimal values from the infimum in
(\ref{orbital-stab}).

Unfortunately, unless $\| q \|_{L^1}$ is assumed to be small, we
cannot prove the analogue of Lemma~\ref{lem:Lax1,0} under the
assumption of small $\| q \|_{L^{\infty}}$. The best we can do is the
bound
\begin{gather*}
\|\varphi_1-1\|_{L^\infty}+\|\varphi_2\|_{L^2} \le C\|q\|_{L^2}, \quad \| \varphi_2 \|_{L^{\infty}} \leq C \| q \|_{L^{\infty}},\\
\|\chi_1\|_{L^2}+\|\chi_2+1\|_{L^\infty}\le C\|q\|_{L^2}, \quad \| \chi_1 \|_{L^{\infty}} \leq C \| q \|_{L^{\infty}}.
\end{gather*}
This is good to control $\| Q - Q_1 \|_{L^{\infty}((-\infty,-x_0) \cup
  (x_0,\infty))}$ in terms of $\| q \|_{L^{\infty}}$ for sufficiently
large $x_0 > 0$ but it is not sufficient to control the $L^{\infty}$-norm
over $(-x_0,x_0)$. More detailed analysis near the soliton core is
needed and the asymptotic stability of $1$-solitons in the cubic NLS
equation is left as an open problem.

{\bf 2.} Another interesting development is a connection between the
NLS equation and the integrable Landau-Lifshitz model
\begin{equation}
\label{LL}
{\bf u}_t = {\bf u} \times {\bf u}_{xx}, \tag{LL}
\end{equation}
where ${\bf u}(t,x) : \R \times \R \to \mathbb{S}^2$ such that ${\bf
  u} \cdot {\bf u} = 1$.
A B\"acklund transformation which connects (\ref{NLS-integrable}) and
(\ref{LL}) is called the Hasimoto transformation (\cite{Has}, \cite{ZT}).
The Hasimoto transformation can potentially be useful to
deduce $L^2$-orbital stability of $1$-solitons of
(\ref{NLS-integrable}) from $H^1$-orbital stability of the domain wall
solutions of (\ref{LL}) and $H^1$-asymptotic stability of $1$-solitons
of (\ref{NLS-integrable}) from $H^2$-asymptotic stability of domain
wall solutions of (\ref{LL}). More studies are needed to see
if our results can be deduced from the corresponding results on
(\ref{LL}) using the Hasimoto transformation.

{\bf 3.} Our approach to employ the B\"{a}cklund transformation for
the proof of $L^2$-orbital stability of solitary waves can be used to
other nonlinear evolution equations integrable by the inverse
scattering transform method. In particular, we expect it to work for
systems where orbital stability of solitary waves in energy space
cannot be deduced by standard methods \cite{GSS}. Nonlinear Dirac
equations in one dimension and Davey-Stewartson equations in two
dimensions are possible examples for applications of our
technique. These examples are left for further studies.


\begin{thebibliography}{16}
\bibitem{Ablowitz} Ablowitz, M.J.; Prinari, B.; Trubach, A.D.;
{\em Discrete and Continuous Nonlinear Schr\"{o}dinger Systems}
(Cambridge University Press, Cambridge, 2004).

\bibitem{BS} V. Buslaev, C. Sulem, On the stability of solitary waves
for Nonlinear Schr\"{o}odinger equations,
Annales Institut Henri Poincar\'e, Analyse Nonlineaire {\bf 202} (2003), 419--475.

\bibitem{CL} T.~Cazenave and P.~L.~Lions,
Orbital stability of standing waves for some nonlinear Schrodinger
 equations, Comm. Math. Phys. \textbf{85} (1982),  549--561.

\bibitem{CW} T. Cazenave and F. Weissler, Rapidly decaying solutions
of the nonlinear Schr\"{o}dinger equation,
Commun. Math. Phys. {\bf 147} (1992), 75--100.

\bibitem{Chen} H.~H.~Chen, General derivation of B\"{a}cklund transformations from inverse scattering problems",
Phys. Rev. Lett. {\bf 33} (1974), 925--928.

\bibitem{I-team}
J.~Colliander, M.~Keel, G.~Staffilani, H.~Takaoka and T.~Tao,
Polynomial upper bounds for the orbital instability of the 1D cubic NLS below
the energy norm, Discrete Contin. Dyn. Syst. \textbf{9} (2003), 31--54.

\bibitem{Cuc} S. Cuccagna, On asymptotic stability in energy space of ground states
of NLS in 1D, J. Diff. Eqs. {\bf 245} (2008), 653--691.

\bibitem{DZ} P. Deift and X. Zhou, Long-time asymptotics for solutions of the NLS equation 
with initial data in a weighted Sobolev space, Comm. Pure Applied Math. {\bf 56} (2003), 1029--1077.

\bibitem{Has} H. Hasimoto, A soliton on a vortex filament,
Journal of Fluid Mechanics \textbf{51} (1972), 477--485.
\bibitem{HN1} N. Hayashi and P.I. Naumkin, Asymptotics for large time of solutions to the nonlinear Schr\"{o}dinger and Hartree equations,
Amer. J. Math. {\bf 120} (1998), 369--389.

\bibitem{HN2} N. Hayashi and P.I. Naumkin, Asymptotic behavior for a quadratic
nonlinear Schr\"{o}dinger equation,
Electron. J. Diff. Eqs. {\bf 2008} (2008), 1--38.

\bibitem{HT} N. Hayashi and M. Tsutsumi, $L^{\infty}(\R^n)$-decay of
classical solutions for nonlinear Schr\"{o}dinger equations,
Proc. Royal Soc. Edinburgh {\bf 104} (1986), 309--327.

\bibitem{Gerard} P. G\'erard and Z. Zhang,
Orbital stability of traveling waves for the one-dimensional Gross-Pitaevskii
equation, J. Math. Pures Appl. {\bf 91} (2009), 178--210.

\bibitem{GV2} J. Ginibre and G. Velo,  On the global Cauchy problem for some
nonlinear Schr\"odinger equations,  Ann. Inst. H. Poincar\'e
Anal. Non Line'aire  \textbf{1}  (1984), 309--323.

\bibitem{GV} J. Ginibre and G. Velo, Scattering theory in the energy space
for a class of nonlinear  Schr\"{o}dinger equations,
J. Math. Pures Appl. {\bf 64} (1985), 363--401.

\bibitem{GSS}
M. Grillakis, J. Shatah, and W. Strauss,  Stability
theory of solitary waves in the presence of symmetry, J. Funct.
Anal., \textbf{74} (1987),  160--197.

\bibitem{Kap} T. Kapitula, On the stability of N-solitons in integrable systems,
Nonlinearity {\bf 20} (2007), 879-907.

\bibitem{Kato} T. Kato, On nonlinear Schr\"odinger equations,
Ann. Inst. H. Poincare' Phys. Th\'eor.  \textbf{46}  (1987), 113--129.

\bibitem{Wadati} K. Konno and M. Wadati,
Simple derivation of B\"{a}cklund transformation from Riccati form of inverse method, Prog. Theor. Phys. {\bf 53} (1975), 1652--1656.

\bibitem{KSF} E.A. Kuznetsov, M.D. Spector, and G.E. Fal'kovich,
On the stability of nonlinear waves in integrable models,
Physica D {\bf 10} (1984), 379--386.

\bibitem{LP} F. Linares and G. Ponce, {\em Introduction to
Nonlinear Dispersive Equations} (Springer, LLC, 2009).

\bibitem{MM2} Y. Martel and F. Merle,
Asymptotic stability of solitons of the subcritical gKdV equations
revisited, Nonlinearity \textbf{18} (2005), 55--80.

\bibitem{MV} F.~Merle, L.~Vega,
$L^2$ stability of solitons for KdV equation,
Int. Math. Res. Not. (2003), 735--753.

\bibitem{Mizum} T. Mizumachi, Asymptotic stability of small solitary waves to
1D nonlinear Schr\"{o}dinger equations with potential, J. Math. Kyoto Univ.
{\bf 48} (2008), 471--497.

\bibitem{MT} T. Mizumachi and N. Tzvetkov,
Stability of the line soliton of the KP-II equation under periodic transverse
perturbations, http://arxiv.org/abs/1008.0812, preprint.

\bibitem{MP} T. Mizumachi and R.~L.~Pego,
Asymptotic stability of Toda lattice solitons, Nonlinearity \textbf{21} (2008),
2099--2111.

\bibitem{O} T. Ozawa, Long range scattering for nonlinear Schr\"{o}dinger equations
in one space dimension,
Commun. Math. Phys. {\bf 139} (1991), 479--493.

\bibitem{T} Y. Tsutsumi, $L^2$-solutions for nonlinear Schr\"{o}dinger equations
and nonlinear groups,  Funkcial. Ekvac.  \textbf{30}  (1987), 115--125.

\bibitem{We}  M. I. Weinstein,
Lyapunov stability of ground states of nonlinear dispersive evolution
equations, Comm. Pure. Appl. Math., \textbf{39} (1986),  51--68.

\bibitem{ZS1} V.E. Zakharov and A.B. Shabat, Exact theory of two-dimensional
self-focusing and one-dimensional self-modulation of waves in nonlinear media,
Soviet Physics JETP {\bf 34} (1972), 62--69.

\bibitem{ZS} V.E. Zakharov and A.B. Shabat,
Interaction between solitons in a stable medium, Soviet Physics JETP {\bf 37}
(1973), 823--828.

\bibitem{ZT} V.E. Zakharov and L.A. Takhtadzhyan,
Equivalence of the nonlinear Schr\"{o}dinger equation and the equation of
a Heisenberg ferromagnet, Theor. Math. Phys. {\bf 38}
(1979), 17--23.
\end{thebibliography}
\end{document}